\documentstyle[psfig,times]{mn}
\newcommand{\araa}{ARA\&A}   
\newcommand{\aj}{AJ}         
\newcommand{\aaa}{A\&A}      
\newcommand{\aas}{A\&AS}     
\newcommand{\aar}{A\&AR}     
\newcommand{\apj}{ApJ}       
\newcommand{\apjs}{ApJS}     
\newcommand{\apss}{Ap\&SS}   
\newcommand{\mnras}{MNRAS}   
\newcommand{\pasp}{PASP}     
 
\newcommand{\Zsun}{$Z_{\odot}$}

\newcommand{\Mzon}{M$_{\odot}$}

\newcommand{\Ha}{H$\alpha$}

\newcommand{\CaII}{Ca\,{\sc ii}}

\def\picplace#1{\vbox{\hrule\@height 0.4pt\@width\hsize
\hbox to\hsize{\vrule\@width 0.4pt\@height#1\hfil
\vrule\@width 0.4pt\@height#1}\hrule\@height 0.4pt\@width\hsize}}
\def\squareforqed{\hbox{\rlap{$\sqcap$}$\sqcup$}}
\def\sq{\ifmmode\squareforqed\else{\unskip\nobreak\hfil
\penalty50\hskip1em\null\nobreak\hfil\squareforqed
\parfillskip=0pt\finalhyphendemerits=0\endgraf}\fi}

\def\la{\mathrel{\mathchoice {\vcenter{\offinterlineskip\halign{\hfil
$\displaystyle##$\hfil\cr<\cr\sim\cr}}}
{\vcenter{\offinterlineskip\halign{\hfil$\textstyle##$\hfil\cr
<\cr\sim\cr}}}
{\vcenter{\offinterlineskip\halign{\hfil$\scriptstyle##$\hfil\cr
<\cr\sim\cr}}}
{\vcenter{\offinterlineskip\halign{\hfil$\scriptscriptstyle##$\hfil\cr
<\cr\sim\cr}}}}}
\def\ga{\mathrel{\mathchoice {\vcenter{\offinterlineskip\halign{\hfil
$\displaystyle##$\hfil\cr>\cr\sim\cr}}}
{\vcenter{\offinterlineskip\halign{\hfil$\textstyle##$\hfil\cr
>\cr\sim\cr}}}
{\vcenter{\offinterlineskip\halign{\hfil$\scriptstyle##$\hfil\cr
>\cr\sim\cr}}}
{\vcenter{\offinterlineskip\halign{\hfil$\scriptscriptstyle##$\hfil\cr
>\cr\sim\cr}}}}}

\def\utw{\smash{\rlap{\lower5pt\hbox{$\sim$}}}}
\def\udtw{\smash{\rlap{\lower6pt\hbox{$\approx$}}}}

\def\fdg{\hbox{$.\!\!^\circ$}}
\def\farcm{\hbox{$.\mkern-4mu^\prime$}}
\def\farcs{\hbox{$.\!\!^{\prime\prime}$}}

\def\diameter{{\ifmmode\mathchoice
{\ooalign{\hfil\hbox{$\displaystyle/$}\hfil\crcr
{\hbox{$\displaystyle\mathchar"20D$}}}}
{\ooalign{\hfil\hbox{$\textstyle/$}\hfil\crcr
{\hbox{$\textstyle\mathchar"20D$}}}}
{\ooalign{\hfil\hbox{$\scriptstyle/$}\hfil\crcr
{\hbox{$\scriptstyle\mathchar"20D$}}}}
{\ooalign{\hfil\hbox{$\scriptscriptstyle/$}\hfil\crcr
{\hbox{$\scriptscriptstyle\mathchar"20D$}}}}
\else{\ooalign{\hfil/\hfil\crcr\mathhexbox20D}}%
\fi}}

\def\bbbc{{\mathchoice {\setbox0=\hbox{$\displaystyle\rm C$}\hbox{\hbox
to0pt{\kern0.4\wd0\vrule height0.9\ht0\hss}\box0}}
{\setbox0=\hbox{$\textstyle\rm C$}\hbox{\hbox
to0pt{\kern0.4\wd0\vrule height0.9\ht0\hss}\box0}}
{\setbox0=\hbox{$\scriptstyle\rm C$}\hbox{\hbox
to0pt{\kern0.4\wd0\vrule height0.9\ht0\hss}\box0}}
{\setbox0=\hbox{$\scriptscriptstyle\rm C$}\hbox{\hbox
to0pt{\kern0.4\wd0\vrule height0.9\ht0\hss}\box0}}}}
\def\bbbq{{\mathchoice {\setbox0=\hbox{$\displaystyle\rm
Q$}\hbox{\raise
0.15\ht0\hbox to0pt{\kern0.4\wd0\vrule height0.8\ht0\hss}\box0}}
{\setbox0=\hbox{$\textstyle\rm Q$}\hbox{\raise
0.15\ht0\hbox to0pt{\kern0.4\wd0\vrule height0.8\ht0\hss}\box0}}
{\setbox0=\hbox{$\scriptstyle\rm Q$}\hbox{\raise
0.15\ht0\hbox to0pt{\kern0.4\wd0\vrule height0.7\ht0\hss}\box0}}
{\setbox0=\hbox{$\scriptscriptstyle\rm Q$}\hbox{\raise
0.15\ht0\hbox to0pt{\kern0.4\wd0\vrule height0.7\ht0\hss}\box0}}}}
\def\bbbt{{\mathchoice {\setbox0=\hbox{$\displaystyle\rm
T$}\hbox{\hbox to0pt{\kern0.3\wd0\vrule height0.9\ht0\hss}\box0}}
{\setbox0=\hbox{$\textstyle\rm T$}\hbox{\hbox
to0pt{\kern0.3\wd0\vrule height0.9\ht0\hss}\box0}}
{\setbox0=\hbox{$\scriptstyle\rm T$}\hbox{\hbox
to0pt{\kern0.3\wd0\vrule height0.9\ht0\hss}\box0}}
{\setbox0=\hbox{$\scriptscriptstyle\rm T$}\hbox{\hbox
to0pt{\kern0.3\wd0\vrule height0.9\ht0\hss}\box0}}}}
\def\bbbs{{\mathchoice
{\setbox0=\hbox{$\displaystyle     \rm S$}\hbox{\raise0.5\ht0\hbox
to0pt{\kern0.35\wd0\vrule height0.45\ht0\hss}\hbox
to0pt{\kern0.55\wd0\vrule height0.5\ht0\hss}\box0}}
{\setbox0=\hbox{$\textstyle        \rm S$}\hbox{\raise0.5\ht0\hbox
to0pt{\kern0.35\wd0\vrule height0.45\ht0\hss}\hbox
to0pt{\kern0.55\wd0\vrule height0.5\ht0\hss}\box0}}
{\setbox0=\hbox{$\scriptstyle      \rm S$}\hbox{\raise0.5\ht0\hbox
to0pt{\kern0.35\wd0\vrule height0.45\ht0\hss}\raise0.05\ht0\hbox
to0pt{\kern0.5\wd0\vrule height0.45\ht0\hss}\box0}}
{\setbox0=\hbox{$\scriptscriptstyle\rm S$}\hbox{\raise0.5\ht0\hbox
to0pt{\kern0.4\wd0\vrule height0.45\ht0\hss}\raise0.05\ht0\hbox
to0pt{\kern0.55\wd0\vrule height0.45\ht0\hss}\box0}}}}
\def\bbbz{{\mathchoice {\hbox{$\sf\textstyle Z\kern-0.4em Z$}}
{\hbox{$\sf\textstyle Z\kern-0.4em Z$}}
{\hbox{$\sf\scriptstyle Z\kern-0.3em Z$}}
{\hbox{$\sf\scriptscriptstyle Z\kern-0.2em Z$}}}}

\newcommand{\BV}{$B\!-\!V$}
\newcommand{\BR}{$B\!-\!R$}
\newcommand{\BI}{$B\!-\!I$}
\newcommand{\BK}{$B\!-\!K$}
\newcommand{\VI}{$V\!-\!I$}
\newcommand{\VK}{$V\!-\!K$}

\newcommand{\IK}{$I\!-\!K$}
\newcommand{\JK}{$J\!-\!K$}
\newcommand{\HK}{$H\!-\!K$}

\begin{document}

\title[Star clusters in NGC 1316]{The
star cluster system of the 3 Gyr old merger remnant NGC~1316: \\
Clues from optical and near-infrared photometry}  

\author[Paul Goudfrooij et al.]{
Paul Goudfrooij,$^{1}$\thanks{E-mail (internet): goudfroo@stsci.edu}
\thanks{Affiliated to the Astrophysics Division, Space Science
Department, European Space Agency}
M.~Victoria Alonso,$^{2}$ 
Claudia Maraston$^{3}$ 
and
Dante Minniti$^{4}$ 
\\ 
$^1$\,Space Telescope Science Institute, 3700 San Martin Drive,
Baltimore, MD 21218, U.S.A. \\
$^2$\,Observatorio Astron\'omico de C\'ordoba and CONICET, Laprida 854, 5000
C\'ordoba, Argentina  \\
$^3$\,Universit\"ats-sternwarte M\"unchen,
Scheinerstrasse 1, D-81679 M\"unchen, Germany \\ 
$^4$\,Department of Astronomy, P.\ Universidad Cat\'olica, Casilla
306, Santiago 22, Chile \\
}

\date{Accepted 2001 July 25. Received ...}

\maketitle

\begin{abstract}
The giant merger remnant galaxy NGC~1316 (Fornax A) is an ideal probe for
studying the long-term effects of a past major merger on star cluster
systems, given its spectroscopically derived merger age of $\sim$\,3 Gyr which
we reported in a recent paper (Goudfrooij et al.\ 2001). Here we present
new ground-based, large-area optical and near-IR imaging of star clusters in
NGC~1316. The ground-based photometry is complemented with deep {\it Hubble
Space Telescope\/} WFPC2 imaging, constituting an excellent combination for
studying globular cluster systems. 
We find that the optical-near-IR colours and luminosities of the brightest
$\sim$\,10 clusters in NGC~1316 are consistent with those of intermediate-age
(2\,--\,3 Gyr) populations. In particular, {\it the near-IR data preclude ages
$\la$ 1.5 Gyr and $\ga$ 4 Gyr\/} for those clusters.  
Unlike `normal' giant ellipticals, the $B\!-\!I$ colour distribution of
clusters in NGC~1316 is not clearly bimodal. However, the luminosity
functions (LFs) of the blue and red parts of the cluster colour distribution
{\it are\/} different: The {\it red\/} cluster LF is well
represented by a power law, $\phi(L)\,dL  
\propto L^{-1.2\pm0.3}\,dL$, extending to about 1.5 mag brighter (in $B$)
than those of typical giant ellipticals.  In contrast, the shape of the {\it
blue\/} cluster LF is consistent with that of `normal' spiral and elliptical
galaxies. 
We conclude that the star cluster system of NGC~1316 is a combination of a
population of age $\sim$\,3 Gyr having roughly solar metallicity and a
population of old, metal-poor clusters which probably belonged to the
pre-merger galaxies. After the 3 Gyr old, metal-rich clusters fade to
an age of 10 Gyr, they will form a red `peak' in a bimodal cluster colour
distribution. This `red peak' will have a colour consistent with that found in
`normal, old' giant ellipticals of the same galaxy luminosity (taking age
dimming into account). 
The surface density profile of clusters in the innermost regions is lower than
that of the integrated light of the galaxy, presumably due to the collective
effect of extended star formation in the inner regions of NGC~1316 and tidal
shocking of the inner clusters. Outside the core, the surface density profile
of clusters is consistent with that of the underlying starlight, suggesting
that the cluster system originally experienced the same violent relaxation as
did the main body of the merger remnant. The specific cluster frequency is
presently $S_N$ = 1.7 $\pm$ 0.4 down to the 50 per cent completeness limit of
the WFPC2 photometry, and will increase to $S_N \ga$ 2.0 as the
merger-induced stellar (and star cluster) population fades to an age of
$\sim$\,10 Gyr (barring further merger events), consistent with specific
frequencies of typical giant ellipticals in the field and in poor groups. 
These features of the star cluster system of NGC~1316 
are fully consistent with 
scenarios for forming `normal' giant elliptical galaxies through gas-rich
mergers  
at look-back times $\ga$ 10 Gyr. 

\end{abstract}



\section{introduction}
\label{s:intro}

Globular clusters (hereafter GCs) 
are among the few observable fossil records of the formation and evolution of
galaxies. Their nature as simple stellar population (hereafter
SSP) significantly simplifies the determination of their ages and
metallicities relative to that of the stellar populations that constitute the
integrated light of their parent galaxies. In our Galaxy, they have yielded
crucial information on early chemical enrichment, the formation time scale of
the galactic halo, the presence of a `thick disk', and an estimate for the
age of the universe. 
While the relative faintness of {\it extragalactic\/} GCs has long prevented
one from obtaining high-quality photometry and (especially) spectroscopy, the
{\it Hubble Space Telescope (HST)\/} and large ground-based telescopes now
allow this goal to be achieved. 

One important, well-known aspect of GC systems among galaxies is that the
number of clusters per unit galaxy luminosity (named the specific frequency
$S_N$) increases systematically from late-type to early-type galaxies, being
$\sim$\,2--3 times higher in elliptical (E) than in spiral galaxies of type
Sb and later (Harris \& van den Bergh 1981; Harris 1991). This fact has been
used as an argument against the scenario for forming ellipticals through
mergers:\ van den Bergh (1995) argued that if ellipticals form through
gas-rich mergers during which star (and GC) formation occurs with a
normal (e.g., Salpeter (1955)) initial mass function (IMF), the $S_N$ of the
resulting merger remnant galaxies should not be significantly different from
the progenitor galaxies since both GCs {\it and\/} stars would have
been formed during the merger. 
On the other hand, recent observations with the {\it HST\/} have led to a
wealth of discoveries of young GCs in merging and starburst galaxies (e.g.,
Holtzman et al.\ 1992; Whitmore et al.\ 1993, 1999; Meurer et al.\ 1995;
Schweizer et al.\ 1996; Miller et al.\ 1997).  
Indeed, the formation of young stars in high-density regions of starbursts
seems to occur preferentially within star {\it clusters\/} rather than
unbound associations (Meurer et al.\ 1995; Schweizer et al.\ 1996), 
most probably from giant molecular clouds whose
collapse is being triggered by a 100-- to 1000-fold increase in gas pressure
due to supernova and shock heating during starbursts (e.g., Jog \& Solomon
1992; Elmegreen \& Efremov 1997). 
As gas-rich galactic mergers produce the most energetic starbursts known
(e.g., Sanders \& Mirabel 1996), the higher  
specific frequency of GCs in ellipticals relative to that in spirals may be
(partly) accounted for by secondary populations of GCs
created during gas-rich mergers. 

In this respect, a particularly interesting feature of the GC 
systems of many giant ellipticals is the presence of bimodal colour
distributions, providing clear evidence for the occurrence of a `second
event' in the formation of these systems (see, e.g., the review by Ashman \&
Zepf 1998 and references therein). While such a bimodal colour distribution
was actually predicted from the merger scenario of galaxy formation (e.g.,
Ashman \& Zepf 1992), opinions about the general nature of the `second
event' differ among authors, as detailed upon below.
One important constraint to
{\it any\/} scenario to explain the bimodality is set by the colour of the
`red' peak in the colour distribution of GCs among giant ellipticals, which
(if interpreted in terms of metallicity) correlates well with the metallicity
and luminosity of the parent galaxy. In contrast, the mean metallicity of the
`blue' peak is more or less constant among galaxies as reported by several
groups (Forbes, Brodie \& Grillmair 1997; Forbes \& Forte 2001; Kundu \&
Whitmore 2001; but see Larsen et al.\ 2001 for a somewhat different view). 
There are three competing scenarios that attempt to explain the bimodality of
GC colours in giant ellipticals. The `merger model' (Schweizer 1987; Ashman
\& Zepf 1992) and the `multi-phase collapse model' (Forbes et al.\ 1997) both
suggest that it is the consequence of GCs forming during two distinct epochs
in the star formation history of these galaxies, whereas the 'accretion
scenario' (Forte, Martinez \& Muzzio 1982; C\^ot\'e, Marzke \& West 1998)
proposes that every galaxy is born with a GC system that has a median colour
(metallicity) according to the colour-magnitude relation among galaxies
(e.g., Terlevich et al.\ 1999), and that the bimodality is due to accretion
of small galaxies (with their associated metal-poor GCs) by a larger galaxy
(with its pre-existing metal-rich GCs). The first two models
differ in the formation mechanism of the metal-rich clusters: the `merger
model' suggests that they are formed during a major, gas-rich galaxy merger,
while the `multi-phase collapse model' proposes that they are formed during a
secondary collapse phase of the galaxy, which does not involve a merger. 

Given these different points of view, it is important to obtain any
additional evidence for (or against) 
the scenario in which the second-generation  GCs
are formed during gas-rich galaxy mergers. 
In this context it is of great interest to find out what the ages
and metallicities of the luminous GCs in merger remnants actually are, and
whether or not the photometric properties of the GC populations found in 
merger remnants will evolve into those typically found in `normal' giant
ellipticals.   
The currently most dependable age determinations stem from Hydrogen Balmer
lines observed in {\it spectra\/} of one bright GC in the peculiar
elliptical galaxy NGC~1275 (Zepf et al.\ 1995a; Brodie et al.\ 1998) and of
two such GCs in NGC~7252 \cite{schsei98}. The luminous GCs in these merger
remnants were found to be $\sim$\,0.5 Gyr old  (cf.\ also 
Hibbard \& Mihos 1995; Schweizer 1998).  
Unfortunately however, the {\it metallicities\/} of these clusters were hard
to measure accurately, since metallic lines are intrinsically weak at this
age (the spectrum being dominated by A-type stars). Thus, it still remained
to be seen whether or not those luminous clusters could evolve into the `red'
clusters seen in `normal' giant ellipticals. 
Another important test of the `merger scenario' is to find out whether or not
these clusters {\it survive\/} the merger era, i.e., whether or not they can
be found in merger remnants at different times after a (major) merger. 
This issue has been brought up for the case of the $\sim$\,0.5 Gyr old
clusters in NGC~1275: Brodie et al.\ (1998) found H$\gamma$ and
H$\delta$ equivalent widths that are somewhat larger (and \BR\ colours
that are bluer) than those of the Bruzual \& Charlot models that they
used, both for Salpeter or Scalo (1986) IMFs. They showed that the
large Balmer line equivalent widths could be brought into agreement
with those models by assuming a truncated IMF involving a mass range
of 2--3 \Mzon. For such a high low-mass cutoff, they argue that these
clusters should fade away in only $\sim$\,1 10$^9$ yr. On the other
hand, the $\sim$\,0.5 Gyr old clusters in NGC~7252 show Balmer line
strengths that are well reproduced by the Bruzual \& Charlot (1996)
models (hereafter the BC96 models) that use the Salpeter or Scalo IMF
(Schweizer \& Seitzer 1998). Also, Gallagher \& Smith (1999) found no
need to invoke a truncated IMF to explain the Balmer line strengths in
young clusters in the starburst galaxy M\,82. But then again, a
detailed study of the very luminous star cluster M\,82--F (Smith \&
Gallagher 2001) revealed that a lower mass cutoff of 2--3 \Mzon\ is
required to match their observations for a Salpeter IMF. Obviously,
the IMF among luminous young star clusters may not be universal.  

We (Goudfrooij et al.\ 2001, hereafter Paper I) recently obtained multi-slit
spectra of GC candidates in NGC~1316 (Fornax A), an early-type merger remnant
galaxy (see, e.g., Schweizer 1980). 
We discovered the presence of $\sim$\,10 GCs associated with NGC~1316 that
have luminosities up to an order of magnitude higher than that of
$\omega$\,Cen, the most massive cluster in our Galaxy. Our measurements of
\Ha\ and the \CaII\ triplet in the spectra of the brightest GCs showed them
to have solar metallicity (to within $\pm$ 0.15 dex) and to have an age of
3.0 $\pm$ 0.5 Gyr. 
This reinforces the view that luminous GCs such as those found before in
younger merger remnants do not necessarily have high low-mass cutoffs to
their IMFs, and it means that they can actually survive disruption processes
taking place during a galactic merger.  

In the present paper we study the photometric properties of NGC~1316's GC
system. We use a combination of {\it HST\/} WFPC2 photometry and
ground-based, large-field optical and near-IR imagery of NGC~1316. The
uniquely high spatial resolution of the WFPC2 data provides accurate 
magnitudes and colours of the GCs, with negligible contamination by
foreground stars or background galaxies. The large-field optical data are used
to determine accurate values for the cluster specific frequency, while the
near-IR data allow us to significantly improve age dating and to search for
intermediate-age, Magellanic-type GCs which feature very red optical-IR
colours, such as found in NGC~5128  \cite{minn+96} and NGC~7252 (Maraston et
al.\ 2001).  

This paper is built up as follows. Section~\ref{s:n1316} summarizes main
properties of NGC~1316. The different observations and data reduction methods
are described in section~\ref{s:obsred}. The various results on the
photometric properties of the GC system are presented in
Section~\ref{s:results}. Section~\ref{s:summ} summarizes 
the main conclusions of this study. Finally, Section \ref{s:outlook} briefly
discusses important outstanding issues. 

\section{NGC 1316}
\label{s:n1316}

\begin{figure*}
\centerline{\psfig{figure=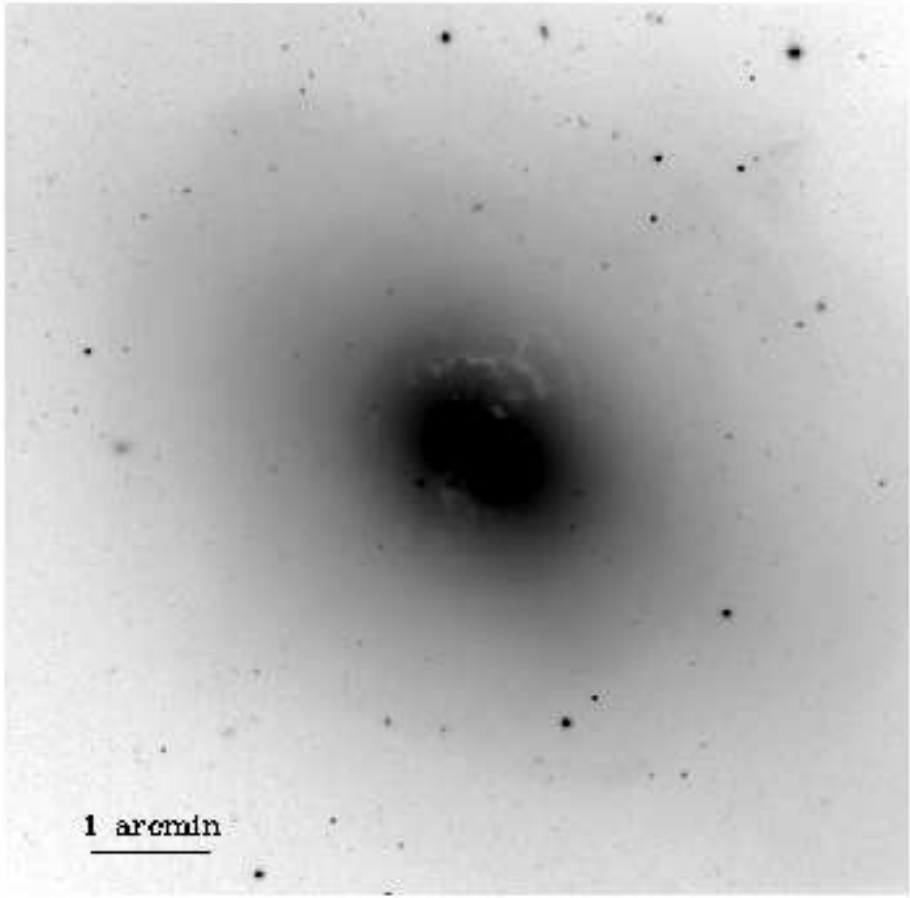,width=12cm}}
\caption[]{Grey-scale plot of $B$-band CCD image of the central 7\farcm5
$\times$ 7\farcm5 of NGC~1316, observed at the NTT (see
Section~\ref{s:NTTobs}). The plate scale is indicated by the bar at the lower
left of the image. North is up and East is to the left.}
\label{f:B_image}
\end{figure*}

The giant early-type galaxy NGC~1316 (= Fornax A = PKS 0320\,$-$\,37 = ESO
357\,$-$\,G022 = ARP 154) is one of the brightest ($B_T$ = 9.42 mag, de
Vaucouleurs et al.\ 1991, hereafter RC3) and closest radio galaxies 
in the sky. It is located in the outskirts of the Fornax cluster, at a
projected distance of 3\fdg7 from NGC~1399, the central giant elliptical
galaxy. Several features of NGC~1316 establish firmly that it is a merger
remnant. The outer envelope includes several non-concentric arcs, tails and
loops that are most likely remnants of tidal perturbations, while the inner
part of the spheroid is characterized by non-concentric ripples 
of 0.1--0.2 mag amplitude above a best-fitting $r^{1/4}$ law, a surprisingly
high central surface brightness and small core radius (and effective radius)
for its galaxy luminosity (Schweizer 1980, 1981; Caon, Capaccioli \&
D'Onofrio 1994). These characteristics, together with a velocity dispersion
which is significantly lower than that of other elliptical galaxies of
similar luminosity, cause NGC~1316 to lie far off the fundamental plane of
early-type galaxies or the Faber-Jackson relation (e.g., D'Onofrio et al.\
1997). Figure \ref{f:B_image} shows a $B$-band image of the
inner 7\farcm5\,$\times$\,7\farcm5 of NGC~1316 in which some of the features
mentioned above can be found. 
All of these features are consistent with NGC~1316 having undergone a
galactic merger after which dynamical relaxation has not yet had time
to complete fully.   

As to the adopted distance of NGC~1316, we take advantage of the fact that two
well-observed type Ia supernovae (SNe\,Ia) have occurred in NGC~1316 (SN1980N
and SN1981D). Using the precise distance indicator for SNe\,Ia that utilizes
the tight relation between their light curve shape, luminosity, and colour
(Riess et al.\ 1998 and private communication), we arrive at a
distance of 22.9 ($\pm$ 0.5) Mpc for NGC~1316, equivalent to $(m\!-\!M)_0$ =
(31.80 $\pm$ 0.05).  At this distance, 1$''$ corresponds to 111 pc. 
Note that this sets NGC~1316 slightly behind the core of the Fornax cluster
(for which $(m\!-\!M)_0$ = 31.54 $\pm$ 0.14; Ferrarese et al.~2000). In fact,
several pieces of evidence suggest that NGC~1316 and a group of smaller
galaxies around it form a dynamical entity that is separate from the main
Fornax cluster: {\it (i)\/} its location in the periphery of the Fornax
cluster, {\it (ii)\/} its luminosity that is 1.1 mag brighter than that of
NGC~1399, the brightest giant elliptical in the Fornax cluster core, and {\it
(iii)\/} the clustering of the radial velocities of galaxies near NGC~1316,
with a median radial velocity that is $\sim$ 300 km s$^{-1}$ higher than that
of the central regions of the Fornax cluster (e.g., Ferguson 1989; RC3). The
dynamical study of the Fornax cluster by Drinkwater, Gregg \& Colless (2001)
indeed shows that the Fornax system has two components: the main Fornax
Cluster centered on NGC 1399 and a small subcluster centered on NGC
1316. Drinkwater et al.\ show that this partition is preferred over a single
cluster at the 99\% confidence level. We therefore assume in the
following that the environment of NGC~1316 is a poor group rather than a rich
cluster. Other global properties of NGC~1316 are listed in  
Table~1 of Paper I. 


\section{Observations, data reduction and cluster photometry}
\label{s:obsred}

\subsection{HST WFPC2 imaging}
\label{s:WFPC2obs}

\subsubsection{Observations and data reduction} 

We retrieved archival images of NGC~1316 taken with the Wide Field
and Planetary Camera 2 (WFPC2) aboard {\it HST\/} in order to study the
star clusters in the inner regions of this galaxy. The
data consist of multiple images through the F450W and F814W
filters. The total exposure times were 5000s in F450W and 1860s in
F814W. 
A subset of the images were spatially offset by 0\farcs5 from
the others, which corresponds to an approximately integer pixel shift
in both PC  and WF CCDs. After standard pipeline processing and
alignment of the images, we combined the images using the STSDAS task
{\sc crrej}, which effectively removed both cosmic rays and hot
pixels. We also trimmed each image to exclude the obscured regions
near the pyramid edges of WFPC2, yielding 751\,$\times$\,751 usable
pixels per CCD.  Results from the authors of the {\it HST\/}
program (proposal ID 5990) have been published in Forbes et al.\
\shortcite{forb+98}, Elson et al.\ \shortcite{elso+98} and Grillmair
et al.\ \shortcite{gril+99}. 

One of the main sources of uncertainty in WFPC2 photometry is
due to charge-transfer efficiency (CTE) problems of the CCDs for which
correction recipes are available (Holtzman et al.\ 1995; Whitmore, Heyer \&
Casertano 1999a). In the present case however, compact-source photometry on a
high and strongly varying background from NGC~1316, we ignored any such
correction. The lowest background surface brightness was encountered in the
F450W WF images, but even there all photometered objects lie in regions where
the background was in excess of 10 counts per pixel. The lack of CTE
correction is therefore expected to affect the magnitudes by less than
0.02 mag and the colours by less than 0.01 mag, negligible for the
purposes of this paper. 

Prior to performing source photometry, the strongly varying
galaxy background was fitted. The main reason for this is to minimize
errors in the photometry due to any particular choice of object
aperture and sky  annulus (see below). For the WF images, the smooth
background gradient was fitted by a bi-cubic spline fit and
subtracted. Care was taken not to include any point-like source or
dust filament in the fit. For the PC image, any fit to the galaxy
background is hampered by the presence of very prominent dust features
reaching into the nucleus (cf.\ Fig.\ 
\ref{f:B_image}). After extensive experimentation, we found 
the following procedure to give the best results. We first built an
isophotal model composed of pure ellipses (separately for each filter)
using the {\sc iraf/stsdas} task {\sc ellipse} in the {\sc isophote}
package. Care was taken to mask out all pixels covered by the dust
features prior to the fitting procedure, as well as bright 
sources, using judgement by eye. Subtraction of the isophotal model
revealed a `residual' image on which compact objects as well as the
dust features stand out. The galaxy background in the parts of
the PC image that were not covered by the elliptical model fit was
fit by applying a median filter with a 11\,$\times$\,11 pixel kernel
to the PC image. After setting the intensity of the median-filtered
image to zero in the locations where the isophotal model was non-zero, the
final `model image' for the PC was created by summing the isophotal
model and the median-filtered image together. 

\subsubsection{Cluster candidate selection and photometry}
 
The selection and photometry of star cluster candidates was carried out
using the {\sc iraf} version of {\sc daophot-ii} \cite{stet87}. The objects
were selected by applying the {\sc daofind} task to an image prepared by
dividing the F450W image by the square root of the model image (thus
having uniform shot noise characteristics over the whole image). We
adopted fairly tight shape constraints [$-$0.6 $<$ {\it roundness}
$<$ 0.6; 0.2 $<$ {\it sharpness} $<$ 0.9] in order to exclude extended
background galaxies and faint objects distorted by noise or
any residual bad pixels. The detection threshold was set at 4 sigma
above the residual background. Although {\sc daofind} returned with
apparent point-like detections located 
within --\,or on the edge of\,-- the dust features, we decided to exclude
those from further analysis due to the difficulty in judging the location of
those sources relative to the dust features along the line of sight,
which is bound to yield spurious results.  

In order to convert the STMAG magnitudes F450W and F814W into 
ground-based Johnson $B$ and Cousins $I$ magnitudes, we used the {\sc synphot}
package within STSDAS. Synthetic spectra of stellar types ranging from B0V to
K3III from the Bruzual \& Charlot \shortcite{brucha93} library were convolved
with Johnson $B$ and Cousins $I$ filters to yield magnitudes relative to Vega,
and with F450W and F814W filters to yield magnitudes in the STMAG
system. Polynomial fits between the two different magnitude systems (to second
order in (F450W\,$-$\,F814W)) resulted in the following conversions: 
\begin{eqnarray}
B-\mbox{F450W} & = & (0.643 \pm 0.002) \nonumber \\
 &  & \quad \mbox{} + (0.138 \pm 0.005)\,(\mbox{F450W}-\mbox{F814W})
 \nonumber \\
 &  & \quad \mbox{} + (0.024 \pm 0.002)\,(\mbox{F450W}-\mbox{F814W})^2 
 \nonumber \\
I-\mbox{F814W} & = & (-1.267 \pm 0.001) \nonumber \\
 &  & \quad \mbox{} + (0.016 \pm 0.003)\,(\mbox{F450W}-\mbox{F814W})
 \nonumber \\ 
 &  & \quad \mbox{} + (0.011 \pm 0.002)\,(\mbox{F450W}-\mbox{F814W})^2 
 \nonumber 
\end{eqnarray}

\begin{figure}
\centerline{\psfig{figure=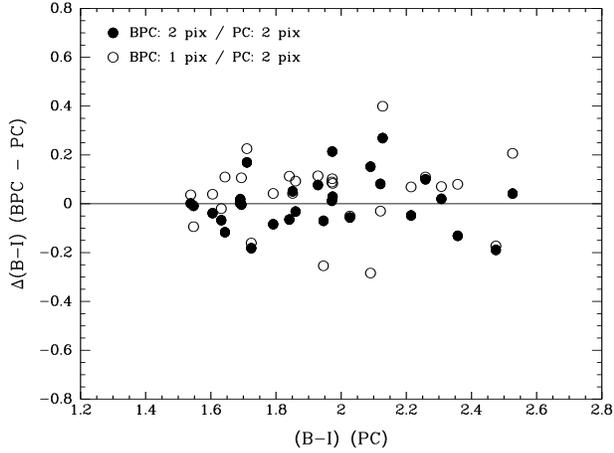,width=8cm,angle=-90.}}
\caption[]{Difference between $B\!-\!I$ colour indices of point-like
sources measured on the unbinned PC CCD (`PC' on the plot) and those
measured on the PC in 2\,$\times$\,2 binned form (`BPC' on the
plot). Solid circles are measurements on the binned PC using an 
aperture radius of 2 (binned) pixels, and open cicles are measurements
on the binned PC using an aperture radius of 1 pixel. The
measurements on the unbinned PC were made with an aperture radius of 2
pixels. Note the relative insensitivity of the photometric accuracy to
the choice of the aperture radius.}
\label{f:apertest}
\end{figure}

Aperture photometry was carried out through small apertures in order
to reduce the noise from the residual background. The background
level was determined in an annulus surrounding each object aperture. After
extensive experimentation with different object aperture sizes, we adopted  
apertures of 2 pixel radius ($r$) for both the PC and WF CCDs.  This choice
was made after considering  
{\it (i)\/} that larger-aperture radii yielded significantly larger errors
for derived $B\!-\!I$ colours, and 
{\it (ii)\/} that smaller aperture radii for the WF CCDs (i.e., 1 pixel)
would be expected to cause systematic biases in the final photometry such as
centering errors and any variation of aperture corrections with position on
the CCD. On the  other hand however, small aperture radii would yield
superior photometric accuracy in the presence of significant variations in
the background. This is why we tested whether a smaller aperture radius (1
WF pixel) for the WF CCDs would improve the photometric accuracy, as
follows. We compared the colours of the point-like sources on the PC with
that of the same sources after having binned the PC CCD by 2\,$\times$\,2
pixels, to simulate the pixel size of the WF CCDs. We performed this
comparison with (a) $r$ = 2 pix for the unbinned PC and $r$ = 1 pix for the
binned PC, and (b) $r$ = 2 pix for the unbinned PC and $r$ = 2 pix for the
binned PC. Aperture corrections were performed according to Holtzman et al.\
\shortcite{holt+95a}. We depict the results of this comparison in Fig.\ 
\ref{f:apertest}.  It can be seen that the accuracy of the photometry is not
very sensitive to the choice of aperture radius. The standard deviation of
$\Delta(B\!-\!I)$, the difference of the colours from the binned and unbinned
PC, was 0.14 for case (a) described above, and 0.11 for case (b). This
encouraging result is mainly due to the successful removal of the galaxy
background as described in the previous subsection. 

Galactic foreground extinction towards NGC~1316 
is $A_B = 0.0$ \cite{burhei84}.

\begin{figure}
\centerline{\psfig{figure=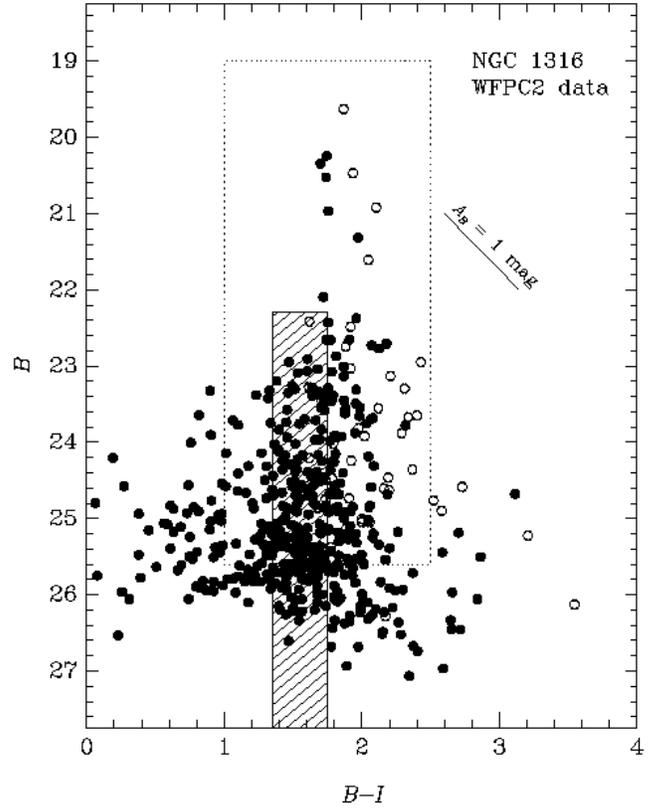,width=8.4cm}}
\caption[]{$B$ vs.\ \BI\ colour-magnitude diagram for NGC~1316 
GC candidates from the WFPC2 data. Only objects with \BI\ errors less than
0.5 mag are plotted. Objects in the PC frame are shown as
open circles, while objects in the WF frames are shown as solid circles. 
The dotted lines delineate the colour and magnitude
criteria for cluster candidates, while the hatched region represents the
apparent magnitude and colour range for GCs in the halo of our Galaxy,
placed at the distance of NGC~1316 (see text for details). 
The solid line shows the reddening vector for $A_B = 1$ mag of extinction.}
\label{f:B_BmI_WFPC2}
\end{figure}

A $B$ vs.\ \BI\ colour-magnitude diagram (hereafter CMD) for the 492 detected
objects is shown 
in Fig.\ \ref{f:B_BmI_WFPC2}, in which the objects in the PC frame are
assigned a symbol different from those in the WF frames. 
The direction of the reddening vector is also shown. Note that the objects
in the PC frame are typically somewhat redder than those in the WF frames,
as could be expected due to the extensive dust filaments in the PC frame. 
From this list of detections, a strict cluster candidate list was made by
applying a colour selection cut of 1.0~$< B\!-\!I <$~2.5, which is the full
range expected for clusters older than 1 Gyr and $-$2.5 $< [Z/Z_{\odot}] <$
0.5 (BC96; Maraston 1998; see also
Section~\ref{s:optpops} below). The quoted lower limit to the age of the
clusters can be considered 
realistic since we found previously from spectroscopy that even the brightest
GCs in NGC~1316 have an age of $\sim$\,3 Gyr (Paper I). For objects in the PC
frame, the upper limit to \BI\ was chosen to be 3.0, allowing for $A_B \la
1$ mag of extinction. The CMD shown in Fig.~\ref{f:B_BmI_WFPC2} is further
discussed in Section~\ref{s:optpops}.  
Photometry and astrometry for the 50 brightest cluster candidates in
the WFPC2 frames is shown in Table~\ref{t:WFPC2phot} in the Appendix. 
A similar table for the full list of cluster candidates in
the WFPC2 frames can be obtained in electronic form from CDS.  

\begin{figure}
\centerline{\psfig{figure=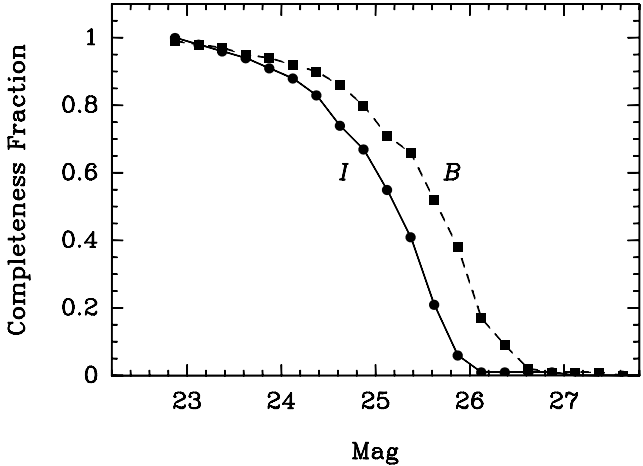,width=8.4cm,angle=-90.}}
\caption[]{$B$ and $I$-band completeness function for the WFPC2 photometry of
NGC~1316 cluster candidates on the WF frames. The 50 per cent completeness
level is at $B \sim$ 25.6 and $I \sim$ 25.2 mag.} 
\label{f:WFcompleteness} 
\end{figure}

Completeness tests were done by adding artificial GCs to the
images and re-applying the aperture photometry programs in a way identical to
those used for the original images. Artificial GCs were 
generated from a composite Point Spread Function (PSF) derived from real GC
candidates in each frame, and added in batches of 100 at different magnitude
intervals, at random positions within the frames, with colours equal to the
median colour of GC candidates. The resulting completeness function for the
WF frames is shown in Fig.\ \ref{f:WFcompleteness}. The 50 per cent
completeness level is at $B \sim 25.6$ for the WF frames, and $B \sim 24.5$
for the PC 
frame. 

\subsection{Ground-based optical imaging}
\label{s:NTTobs}

\subsubsection{Observations and data reduction}

Ground-based CCD images of NGC~1316 were retrieved from the ESO archive
database. The data were originally obtained 
in November 1992 
at the ESO 3.5-m New Technology Telescope
(NTT) with the red arm of the ESO Multi-Mode Instrument (EMMI). The detector
was a thick, front-illuminated CCD of type Thomson 31156 Grade A, having
1024$\times$1024 sensitive pixels. The pixel size was 
$19\mu\mbox{m}\!\times\!19\mu$m, yielding a scale of 0.44 arcsec per
pixel and a total field of view of 7\farcm5\,$\times$\,7\farcm5 (see 
Fig.\ \ref{f:B_image}).  NGC~1316 was
observed through $B$, $V$ and $I_C$ (Cousins) filters, with total exposure
times of 4200 s, 2100 s and 900 s, respectively, split up into several
exposures.

After retrieving several bias, dark, and twilight flat field frames taken
during the observing runs in which the images of NGC~1316 were taken,
basic data reduction was carried out 
in the standard way using the {\sc iraf ccdred} package. Cosmic radiation
events in the individual images were removed by an appropriate averaging
program. First, the images taken through the same filter were aligned to a
common coordinate system, using the centroids of stars in the field of
view. This alignment procedure was accurate to within 0.03 pixel. The images
were subsequently averaged together by comparing all individual pixel values
(per unit exposure time) with the median value over all frames (taken using
the same filter). Individual pixels are rejected during the combining process if 
their value exceeds the range expected from the (sky + read-out) noise. 
The effective seeing values of the combined images are 1\farcs28 in $B$,
1\farcs23 in $V$, and 1\farcs17 in $I_C$. Fig.~\ref{f:B_image} shows the
final combined $B$-band image. 

Photometric calibrations were derived using images of the 
Rubin~149 standard star field for which the photometry in the
Johnson--Kron--Cousins system is given in Landolt \shortcite{land92}. The
resulting colour equations were:  
\begin{eqnarray}
B-b & = & (22.977 \pm 0.002) + (0.032 \pm 0.002)\,(b\!-\!v) \nonumber \\
V-v & = & (24.466 \pm 0.005) + (0.027 \pm 0.005)\,(b\!-\!v) \nonumber \\
I-i & = & (24.060 \pm 0.002) - (0.020 \pm 0.003)\,(v\!-\!i) \nonumber 
\end{eqnarray}
where upper case $BVI$ denote calibrated Johnson--Kron--Cousins magnitudes,
and lower case $bvi$ denote instrumental magnitudes (normalized to an exposure
time of 1 s). We used the standard 
atmospheric extinction coefficients for the ESO site (ESO User Manual,
1993). This calibration was double checked with WFPC2 photometry of
stars in the area covered by the WFPC2 images and with published aperture
photometry of Wegner \shortcite{wegn79}. The differences were minor (less than
0.02 mag in any band). 

\subsubsection{Cluster candidate selection and photometry}

As described above for the case of the {\it HST\/} WFPC2 images, we attempted
to remove the galaxy background from the images. First we used 
the STSDAS task {\sc ellipse} to perform isophotal model fits to the 
surface brightness distribution of the galaxy (after masking out all pixels
covered by dust features and bright sources). Subtraction of that model
revealed a rich system of extensive shells and ripples (cf.\ also Schweizer
1980), as well as the dust features, small galaxies and compact sources. To
also remove the shells and ripples, we first subtracted all compact sources
using the {\sc daophot-ii} PSF fitting task. Then we applied a median filter
(kernel = 9\,$\times$\,9 pixels) to the image. Finally, the sum of the
isophotal model and the median-filtered image constituted the final model
image. 

The sky background levels and their uncertainties were determined by fitting
power laws to the outer parts of the intensity profiles of the isophotal
model as described in Goudfrooij et al.\ \shortcite{paul+94a}. 
The selection and photometry of star cluster candidates was carried out using
{\sc daophot-ii}. After aligning the $B$, $V$ and $I_C$ images on the same
coordinate system, the {\sc daofind} task was run on the $V$ image (again,
divided by the square root of the model image) which had the
highest signal-to-noise. 
The detection threshold was set at 4 sigma above the
(residual) background. Due to the very complex dusty structure of the
central regions of NGC~1316, especially at ground-based resolution, we
decided to exclude the compact sources found within the central
$\sim$\,50$''$ from the nucleus from further analysis. The great majority of
those compact sources was measured on the WFPC2 images anyway. 

\begin{figure}
\centerline{\psfig{figure=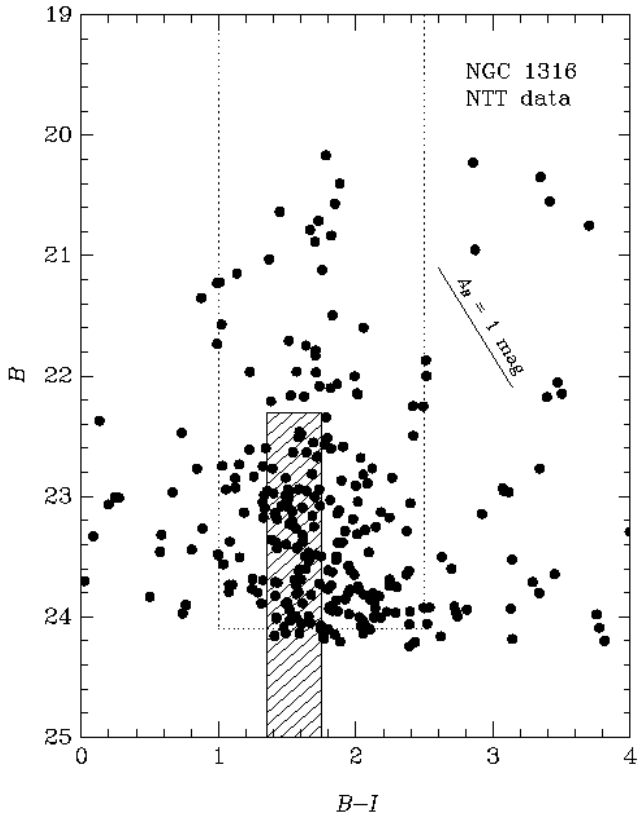,width=8.4cm}}
\caption[]{$B$ vs.\ \BI\ colour-magnitude diagram for NGC~1316 
GC candidates from the NTT data. Only objects with \BI\ errors less than
1.0 mag are plotted. The dotted lines delineate the colour and magnitude
criteria for cluster candidates, while the hatched region represents the
apparent magnitude and colour range for GCs in the halo of our Galaxy,
placed at the distance of NGC~1316 (see text for details). 
The solid line shows the reddening vector for $A_B = 1$ mag of extinction.}
\label{f:B_BmI_NTT}
\end{figure}

Aperture photometry of the cluster candidates was performed using an aperture
radius of 4 pixels. Aperture corrections were derived for each filter from a
curve-of-growth analysis of 10 well-exposed, isolated stars in the field. 

\begin{figure}
\centerline{\psfig{figure=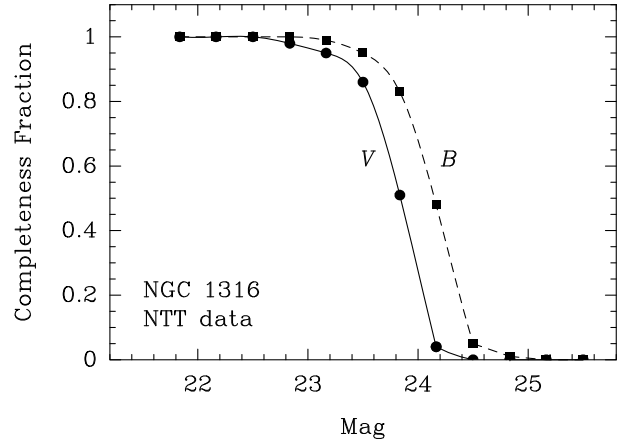,width=8.cm,angle=-90.}}
\caption[]{$B$ and $V$ band completeness functions for the NTT photometry of
NGC~1316 cluster candidates. The 50 per cent completeness levels are at $B
\sim 24.10$ and $V \sim 23.85$.} 
\label{f:NTTcompleteness}
\end{figure}

A $B$ vs.\ \BI\ CMD for the 350 detected objects is
shown in Fig.\ \ref{f:B_BmI_NTT}. The direction of the reddening vector is
also shown. From this list of detections, a strict cluster candidate list was
made by applying a colour selection criterion of $1.0 < B\!-\!I < 2.5$,
i.e., the same as for the WF frames of the {\it HST\/}
data. Fig.~\ref{f:B_BmI_NTT} is further discussed in
Section~\ref{s:optpops}. 
Photometry and astrometry for the 50 brightest cluster candidates in
the NTT frames is shown in Table~\ref{t:NTTphot} in the Appendix. 
A similar table for the full list of cluster candidates in
the NTT frames can be obtained in electronic form from CDS.  

Completeness tests were done the same way as for the {\it HST}/WFPC2 images
(cf.\ Section \ref{s:WFPC2obs}). The resulting completeness functions for
the $B$ and $V$ images are shown in Fig.\ \ref{f:NTTcompleteness}. The
50 per cent completeness levels are at $B \sim 24.10$ and $V \sim
23.85$. These represent our faint magnitude cut-offs for the cluster
candidates. Even using these selection criteria we do expect some
contamination by foreground stars and background galaxies in the cluster
candidate sample, due to the lower spatial resolution with respect to the
{\it HST\/} data. This issue will be discussed below in Section
\ref{s:S_N}.1.  
\subsection{Ground-based near-infrared imaging}
\label{s:IRobs}

\subsubsection{Observations and data reduction}

Near-infrared images of NGC~1316 
were obtained during the nights of 1996 November 22--25 using the
IRAC2 camera mounted on the ESO/MPI 2.2-m telescope. The detector was
a Rockwell NICMOS3 array (HgCdTe, 256\,$\times$\,256 pixels). We used
lens C, having an image scale of 0\farcs509 per pixel. The read-out
noise was 22.5 e$^-$ per pixel.  We mapped a region of $\sim$\,19
arcmin$^2$ around the centre of NGC~1316 with
the $J$, $H$ and $K'$ filters. The field covering the central region of
NGC~1316 was exposed for a longer time than the other fields in order to get
the highest possible signal-to-noise ratio in the region overlapping the WFPC2
field. The sky was photometric during the whole observing run, and the seeing
varied between 0\farcs8 and 1\farcs0. Separate sky frames located at
$\ga 20'$ from the galaxy centre were also taken, interleaved with the
individual galaxy frames to enable proper sky subtraction. 
The exposure times for each individual frame were chosen to be as long as
possible, but care was taken to have the intensity at the galaxy centre stay
within the 1 per cent linearity regime. 

The reduction procedure was as follows. First, all dark frames with identical
integration time were combined and subtracted from the raw images (i.e., dome
flats, standard star frames, sky frames, and galaxy frames). For each filter,
the dome flats taken with identical lamp voltages were combined, and the
combined zero-voltage dome flat was subtracted from all other dome flats. We
used a combined dome flat with an intensity level similar to that of the
object frames for final flat fielding, after having normalized it to a mean
value of unity. The dome flats were also used to create a bad pixel mask which
allowed the replacement of bad pixels by the median of the values of the
surrounding pixels. A final correction to the low-frequency flat field
correction (`illumination correction') was made by mapping a standard star
on a grid of 16 positions across the surface of the array, and fitting a
third-order polynomial to the total intensities of the star at the different 
positions.  

The photometric calibration was derived from observations of several standard
stars from Carter \& Meadows \shortcite{carmea95}. Each standard star
was observed on 2 regions of the array that are free of bad pixels. Aperture
photometry was then performed on an image that is the difference of the two
flat-fielded standard star images, ensuring a proper sky subtraction. The
resulting colour equations were:  
\begin{eqnarray}
J-j & = & (21.930 \pm 0.008) \nonumber \\
 &  & \quad \mbox{} + (0.148 \pm 0.014)\,(j\!-\!k') - 0.15\,x \nonumber \\
H-h & = & (21.706 \pm 0.006) \nonumber \\
 &  & \quad \mbox{} + (0.028 \pm 0.010)\,(j\!-\!k') - 0.10\,x \nonumber \\
K-k' & = & (21.170 \pm 0.006) \nonumber \\
 &  & \quad \mbox{} + (0.022 \pm 0.017)\,(j\!-\!k') - 0.17\,x \nonumber 
\end{eqnarray}
where upper case $JHK$ denote calibrated magnitudes, lower case $jhk'$
denote instrumental magnitudes normalized to an exposure time of 1 s, and $x$
is the airmass. These zeropoints were double checked by comparing the $JHK$
aperture photometry of NGC~1316 by Glass \shortcite{glas84}\ with our
data. They turned out to be consistent with one another to within 0.03 mag in
any passband.  

\begin{figure*}
\ \\
\centerline{\psfig{figure=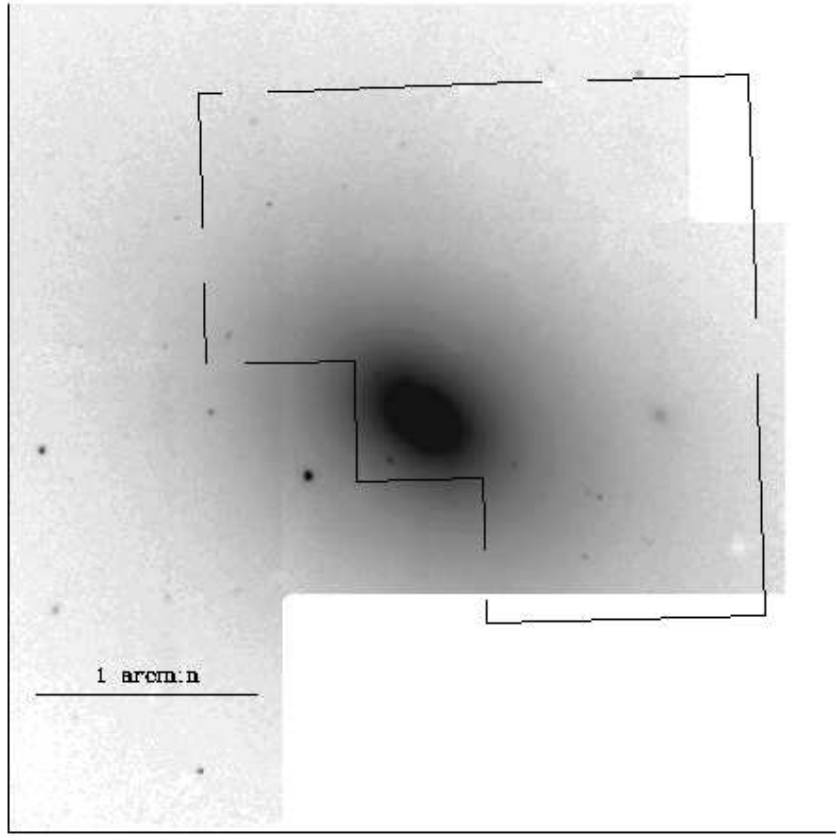,width=11cm,height=11cm}}
\caption[]{Grey-scale plot of $H$-band mosaic of the central 3\farcm75
$\times$ 3\farcm75 of NGC~1316, observed with IRAC2B at the ESO/MPI 2.2-m
telescope (see Section~\ref{s:IRobs}). The location of the WFPC2 frames is
outlined in black. The plate scale is indicated by the bar
in the lower left of the image. North is up and East is to the left. 
The faint, extended `object' located $\sim$\,1 arcmin west of NGC 1316 is a
detector feature (caused by amplifier crosstalk).}
\label{f:H_image}
\end{figure*}

\subsubsection{Cluster candidate selection and photometry}

In order to put the optical and near-IR images on a common coordinate system,
we measured the centroids of several point sources in common between the 
near-IR images and the NTT $V$-band image, using the {\sc iraf} task {\sc
xyxymatch}. The alignment error was always (and, mostly, far) below 0.1
pixel. All near-IR images were subsequently individually transformed to the
NTT coordinate system. Finally, all transformed near-IR images were mosaiced
together per filter. Fig.\ \ref{f:H_image} shows the final $H$-band mosaiced
image, with the location of the WFPC2 field superposed onto it. 
The galaxy background subtraction, point-source photometry and the derivation
of aperture corrections were performed in the same way as for the optical NTT
images. The near-IR photometry of the cluster candidates 
are listed in Tables~\ref{t:WFPC2phot} and \ref{t:NTTphot} in Appendix A.   


\section{Properties of the star cluster system of NGC~1316}
\label{s:results}

\subsection{Optical properties of stellar populations}
\label{s:optpops}

\begin{figure}
\centerline{\psfig{figure=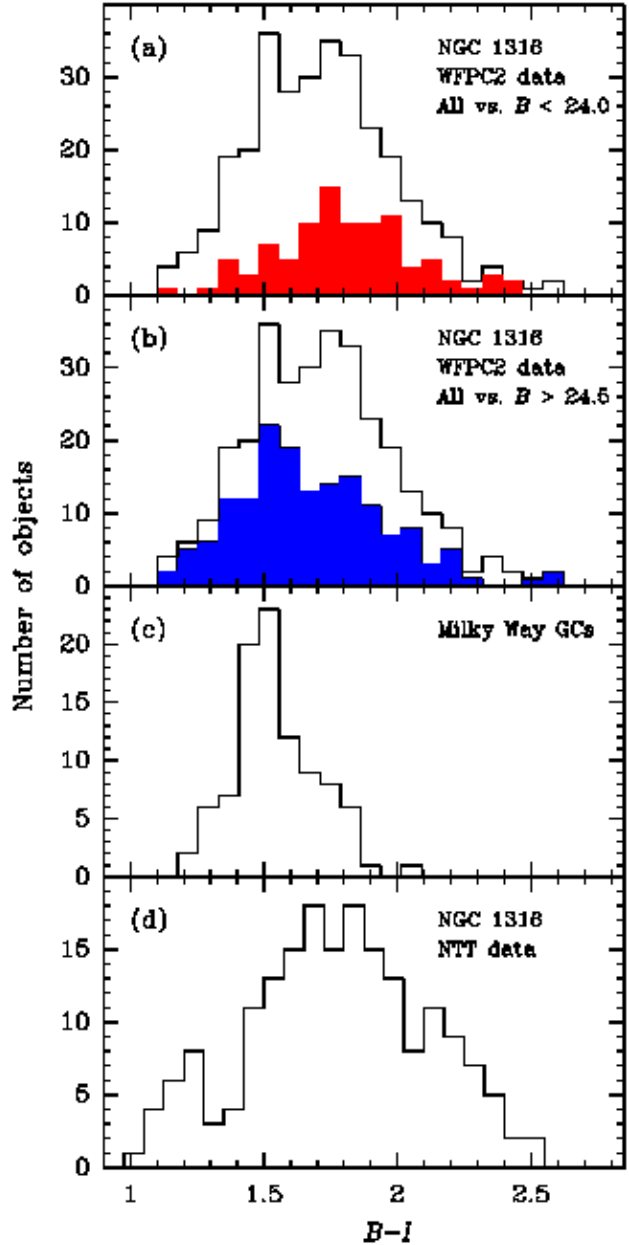,width=8.1cm}}
\caption[]{Histograms of the \BI\ colours of NGC~1316 cluster candidates 
compared with those of GCs in the Milky Way. In 
panels {\bf (a)} and {\bf (b)}, the open histogram depicts the colour
distribution of {\it all\/} cluster candidates in the WFPC2 field of
NGC~1316. The shaded histogram in panel {\bf (a)} depicts the colour
distribution of the NGC~1316 cluster candidates with $B < 24.0$ in the WFPC2
photometry, while the shaded histogram in panel {\bf (b)} depicts
the colour distribution of the NGC~1316 cluster candidates with $B >
\mbox{24.5}$ in the WFPC2 photometry. Panel {\bf (c)} depicts the histogram
of \BI\ colours of 
GCs in the Milky Way. Panel {\bf (d)} depicts the histogram of 
\BI\ colours of NGC~1316 cluster candidates from the NTT photometry.}  
\label{f:BIhists}
\end{figure}

\subsubsection{Interpretation of optical photometry}

The $B$ vs.\ $B-I$ CMD for cluster candidates in the WFPC2
and NTT frames was shown in Figs.~\ref{f:B_BmI_WFPC2} and \ref{f:B_BmI_NTT},
respectively. The area which the GC system of the halo of the Milky Way
would occupy if placed at the distance of NGC~1316 is also indicated, using
the database of Harris (1996). These diagrams suggest some interesting
results.  

First, the \BI\ colours of the clusters that are brighter than the brightest
GC in our Galaxy are quite uniform and on the red end of the range covered by
the Galactic GCs (and beyond). Spectra of three of these bright clusters were
analyzed in Paper I, where we found all three to be $\sim$\,coeval with an
age of 3.0 $\pm$ 0.5 Gyr and a solar metallicity (to within 0.15 dex). The
\BI\ colours of these bright clusters are consistent with their
spectroscopically derived age and metallicity range (Paper I; see also
below).  

Second, the clusters fainter than $B \sim \mbox{23}$ tend to be somewhat
bluer on average than the brighter clusters (excluding some inner clusters in
the PC frame which are likely reddened by dust, cf.\
Fig.~\ref{f:B_BmI_WFPC2}). This trend is more obvious from the WFPC2 diagram 
(Fig.~\ref{f:B_BmI_WFPC2}) than from the NTT diagram
(Fig.~\ref{f:B_BmI_NTT}), which is most likely due to the higher
photometric accuracy of the WFPC2 data relative to that of the NTT data.  
The range of colours encompassed by the clusters with $B \ga \mbox{23}$ is
roughly consistent with that encompassed by the Galactic GCs, considering the
photometric errors (e.g., $\sigma_{B-I} \sim \mbox{0.13}$ at $B \sim
\mbox{25}$). The main cause of the bluer mean colours at fainter brightness
levels would seem to be a lower metallicity rather than a younger age, as
younger populations are brighter than older ones. These fainter clusters may
be expected to be a mixture of {\it (i) old\/}, metal-poor clusters associated
with the pre-merger galaxies (given their overlapping colours and
luminosities) and {\it (ii)\/} the lower-luminosity end of the system of
second-generation clusters formed during the merger. 

These results are illustrated further in
Fig.~\ref{f:BIhists} which shows \BI\ colour distributions of cluster
candidates on the WFPC2 frames (in the form of histograms), and compares them
with that of the GC system of the Milky Way. The colour distribution of the
full cluster sample (depicted as open histogram on panels (a) and (b) of
Fig.~\ref{f:BIhists}) exhibits two peaks: one at $B\!-\!I \sim \mbox{1.5}$ 
{\it [coinciding with the peak of the Milky Way GC system, cf.\ panel (c)]\/}
and one at $B\!-\!I \sim \mbox{1.8}$. Panel (a) shows clearly that the bright
end of the cluster luminosity function (hereafter LF; shown as filled
histogram; many of these objects were {\it proven\/} to be genuine clusters in
Paper I) primarily 
makes up the red 
peak of the overall colour distribution, while the panel (b) shows that at
the faint end of the cluster LF, the blue peak of the colour distribution
becomes about as highly populated as the red peak.  The reddest `wing' to
the colour distribution in Fig.~\ref{f:BIhists} is mostly populated by
cluster candidates in the innermost regions (i.e., the PC frame,
cf.\ Fig.~\ref{f:B_BmI_WFPC2}). This suggests that the most likely main
cause of the red wing is dust extinction rather than high metallicity. To
decide unambiguously between these possibilities, spectroscopy and/or
near-IR photometry with 8\,--\,10\,m-class telescopes will be needed. 
Finally, the colour distribution of the cluster candidates on the NTT frames
(outside a radius of 50$''$, like in Fig.~\ref{f:B_BmI_NTT}) is shown in
panel (d) of Fig.~\ref{f:BIhists}. It is similar to that of the clusters 
on the WFPC2 frames, except that the blue peak is less populated in
the former. This is entirely consistent with our interpretation involving two
cluster populations mentioned above, given the fact that the NTT photometry
doesn't reach deep enough to detect the bulk of the fainter, bluer clusters
that are supposedly the `old' clusters from the pre-merger galaxies. 

\begin{figure*}
\centerline{~~
\psfig{figure=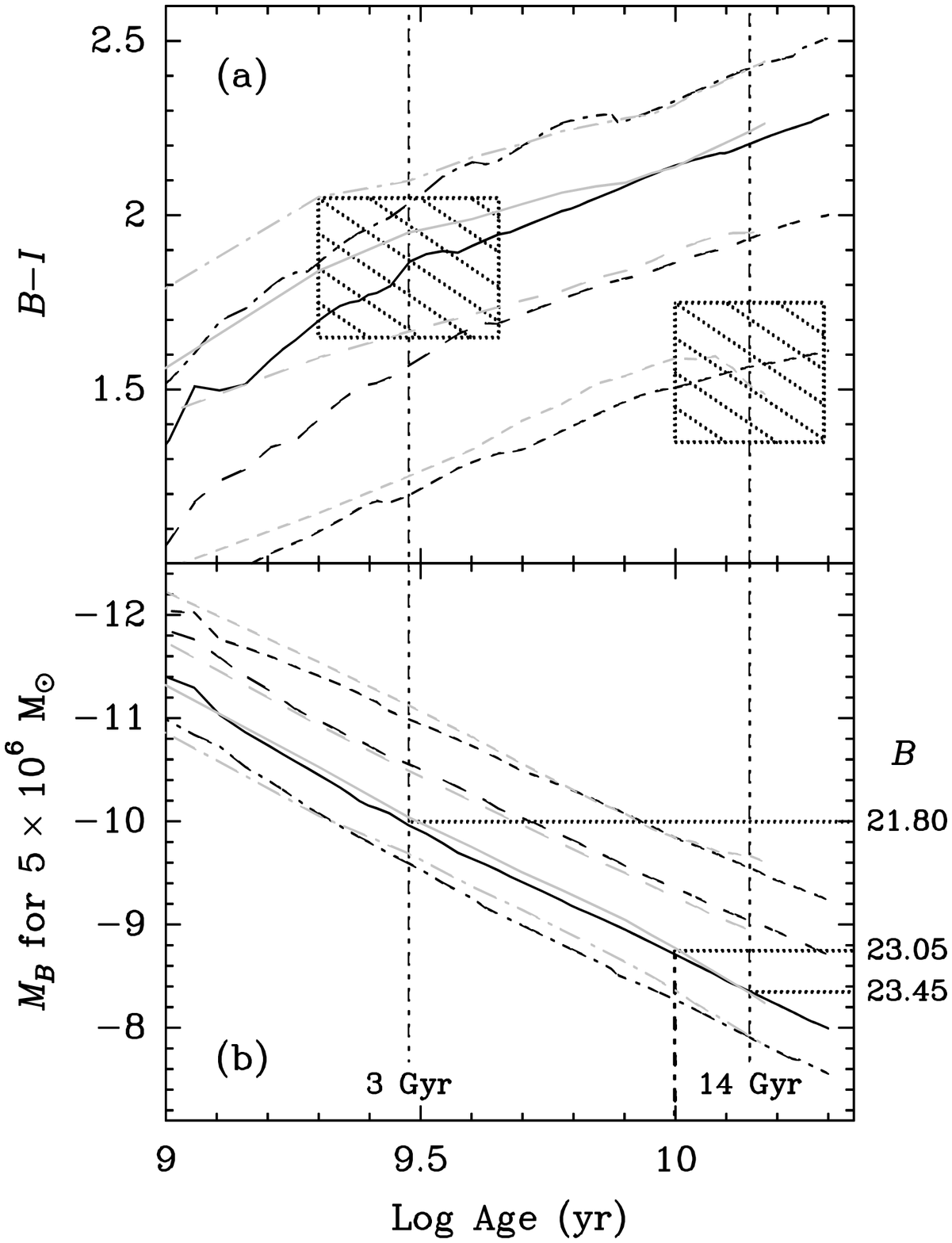,width=8.4cm}
\hfill\
\psfig{figure=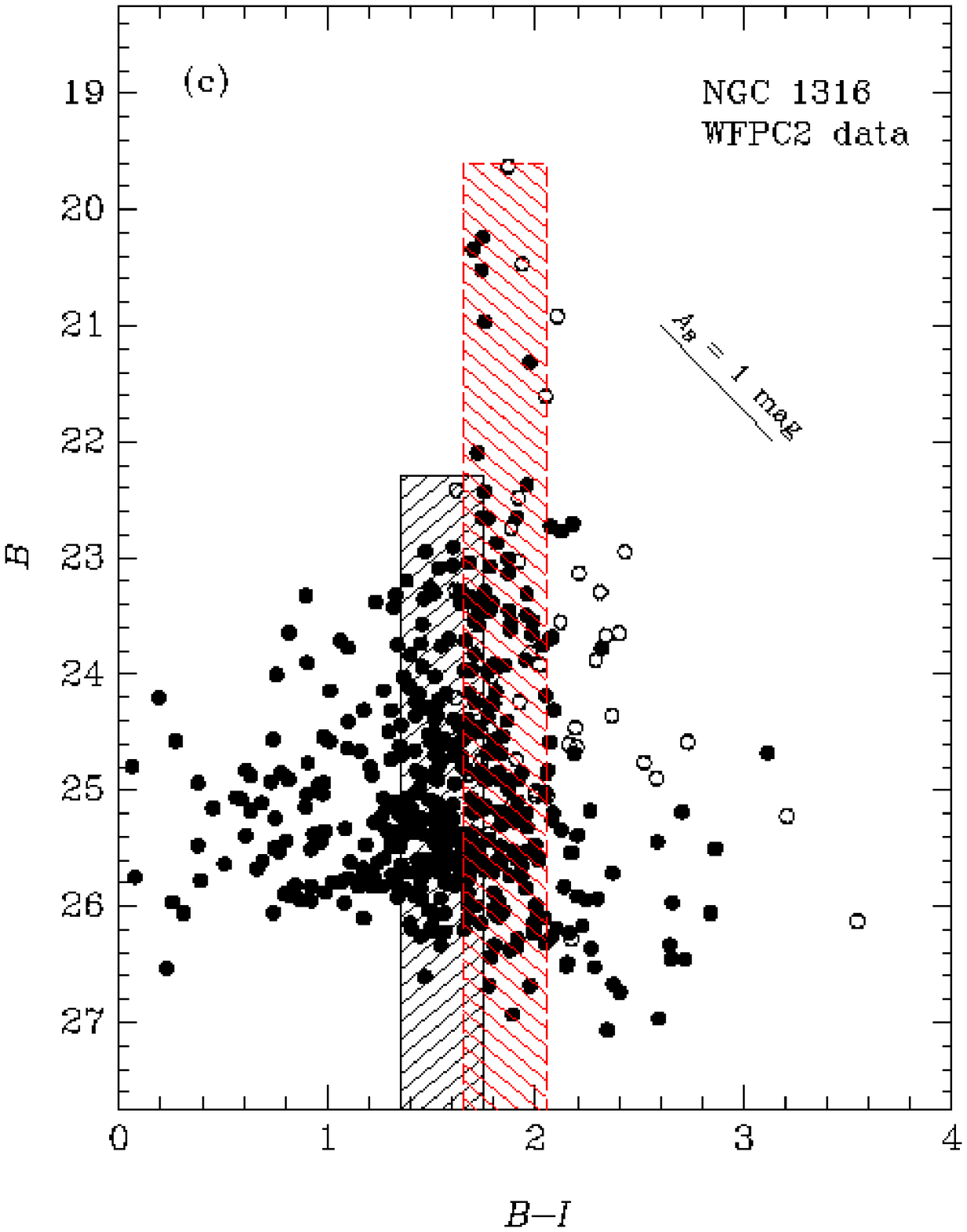,width=8.2cm,height=10.8cm}~~
}
\caption[]{{\sl Panel {\bf (a)}}:\ Time evolution of $B\!-\!I$ colour index
 according to single-burst stellar population models for ages older than 1
 Gyr. The black lines represent the Bruzual \&\ Charlot (1996) models, while
 the grey lines represent the Maraston (2001) models (described in Sect.\
 \ref{s:irpops}).  
 Model curves are plotted for a 
 Salpeter (1955) IMF and the following metallicities:\ 0.02 solar {\it
 (short-dashed lines)}, 0.2 solar {\it (long-dashed lines)}, 1.0 solar {\it
 (solid lines)}, and 2.5 solar {\it (dot-long-dashed lines)}. Ages of 3 Gyr
 and 14 Gyr are indicated as vertical short-long-dashed lines. The region
 hatched by dotted lines on the {\it 
 left\/} indicates the $B\!-\!I$ colour interval populated by 3 Gyr old
 clusters in NGC~1316 with a solar metallicity to within 0.15 dex, while
 the hatched region on the {\it right\/} does so for a old, metal-poor cluster
 population such as that of the Milky Way halo [the same \BI\ intervals are
 indicated in Panel (c)]. \\
 {\sl Panel {\bf (b)}}:\ Evolution of $B$-band magnitudes (labels are absolute
 magnitudes on the left-hand side, apparent magnitudes at the distance of
 NGC~1316 on the right-hand side) for a globular cluster having the mass of
 $\omega$\,Cen (5 $\times$ 10$^6$ \Mzon, Meylan et al.\ 1995). Curves as in
 Panel (a). Ages of 3 Gyr, 10 Gyr, and 14 Gyr are indicated. The horizontal
 dotted lines indicate $B$ magnitudes for specific cases, as discussed in
 Section~\ref{s:optpops}.2. \\ 
 {\sl Panel {\bf (c)}}:\ Same $B$ vs.\ \BI\ colour-magnitude diagram as in
 Fig.~\ref{f:B_BmI_WFPC2}. The hatched region on the {\it left\/} (i.e.,
 1.35~$\le$ \BI\ $\le$~1.75) delineates the apparent magnitude and colour
 range for GCs in the halo of our Galaxy (placed at the distance of
 NGC~1316), while the hatched region on the {\it right\/} (i.e.,
 1.65~$\le$ \BI\ $\le$~2.05) does so for a 3 Gyr old population of
 star clusters in NGC~1316 with a metallicity of [$Z/Z_{\odot}$] =
 0.00 $\pm$ 0.15 dex. See discussion in Section~\ref{s:optpops}.1.
}
\label{f:twopops}
\end{figure*}

Summarizing the optical photometry results presented above, the blue peak of
the overall \BI\ histogram, which coincides with the modus of the \BI\
colour distribution of the Galactic GC system, is primarily populated by
the faint end of the cluster LF, while the red peak in the \BI\ histogram is
solely populated by the bright end of the cluster LF. Since we also know from
spectroscopy that the brightest clusters in NGC~1316 are 3 $\pm$ 0.5 Gyr old
and have a solar metallicity to within 0.15 dex (Paper I), an
obvious interpretation of these results is that we are dealing with the
presence of two distinct cluster populations:\   
{\it (i)\/} a second-generation population clusters of roughly solar
metallicity that were formed $\sim$\,3 Gyr ago during a major spiral-spiral
merger (as determined from spectra in Paper~I), and 
{\it (ii)\/} a population of old, primarily metal-poor clusters that were
associated with the progenitor galaxies. From the colour histograms of the
WFPC2 data shown in panels (a) and (b) of Fig.\ \ref{f:BIhists}, we estimate
that of order 50 per cent of the total number of clusters is of the
second generation. 

Is this interpretation consistent with stellar population model
predictions\,? We considered the following to address this question. 
The metallicity range encompassed by the brightest clusters at an age of
3~$\pm$~0.5 Gyr (derived from spectra in Paper I:\ $-\mbox{0.15} \la
[Z/Z_{\odot}] \la \mbox{0.15}$) corresponds to $\mbox{1.75} \le B\!-\!I \le
\mbox{1.95}$ according to the BC96 models. Considering the typical
photometric errors of the WFPC2 photometry as well (e.g., $\Delta(B\!-\!I)
\la \mbox{0.1}$ for $B \le \mbox{24}$), we assigned a \BI\ interval of
$\mbox{1.65} \le B\!-\!I \le \mbox{2.05}$ to the putative second-generation
population of clusters. This is indicated as the hatched region on the {\it
left\/} side of panel (a) of Fig.~\ref{f:twopops}, which depicts the
predicted time evolution of the \BI\ colour for single-burst populations
older than 1 Gyr [using the BC96 models as well as the Maraston (2001; see
Sect.\ \ref{s:irpops}) models for comparison]. 
The hatched region on the {\it right\/} side of panel (a) indicates the \BI\
interval occupied by the Galactic GC system which corresponds to a
metallicity range of $-\mbox{1.5} \la [Z/Z_{\odot}] \la -\mbox{0.5}$ (Harris
1996; Minniti 1995) which is consistent with the model predictions at an
age of 14 Gyr. 
These two \BI\ intervals are also indicated on panel (c) of
Fig.~\ref{f:twopops}, which is the $B$ vs.\ \BI\ CMD from the WFPC2
photometry (i.e., the same as Fig.~\ref{f:B_BmI_WFPC2}). It can be seen that
these two \BI\ intervals provide a very good fit to the positions of the
cluster candidates on the CMD. The upper brightness limit to the hatched
region for the second-generation population of clusters on panel (c) was
chosen to encompass the brightest cluster candidates, since the spectra in
Paper I showed them to be genuine clusters associated with NGC~1316. Note
that this brightness limit is several magnitudes brighter than that of the
Galactic cluster system, rendering the brightest second-generation clusters
in NGC~1316 to be much more massive than $\omega$\,Cen, the brightest cluster
in our Galaxy (this was discussed in detail in Paper I).  

\begin{figure}
\centerline{\psfig{figure=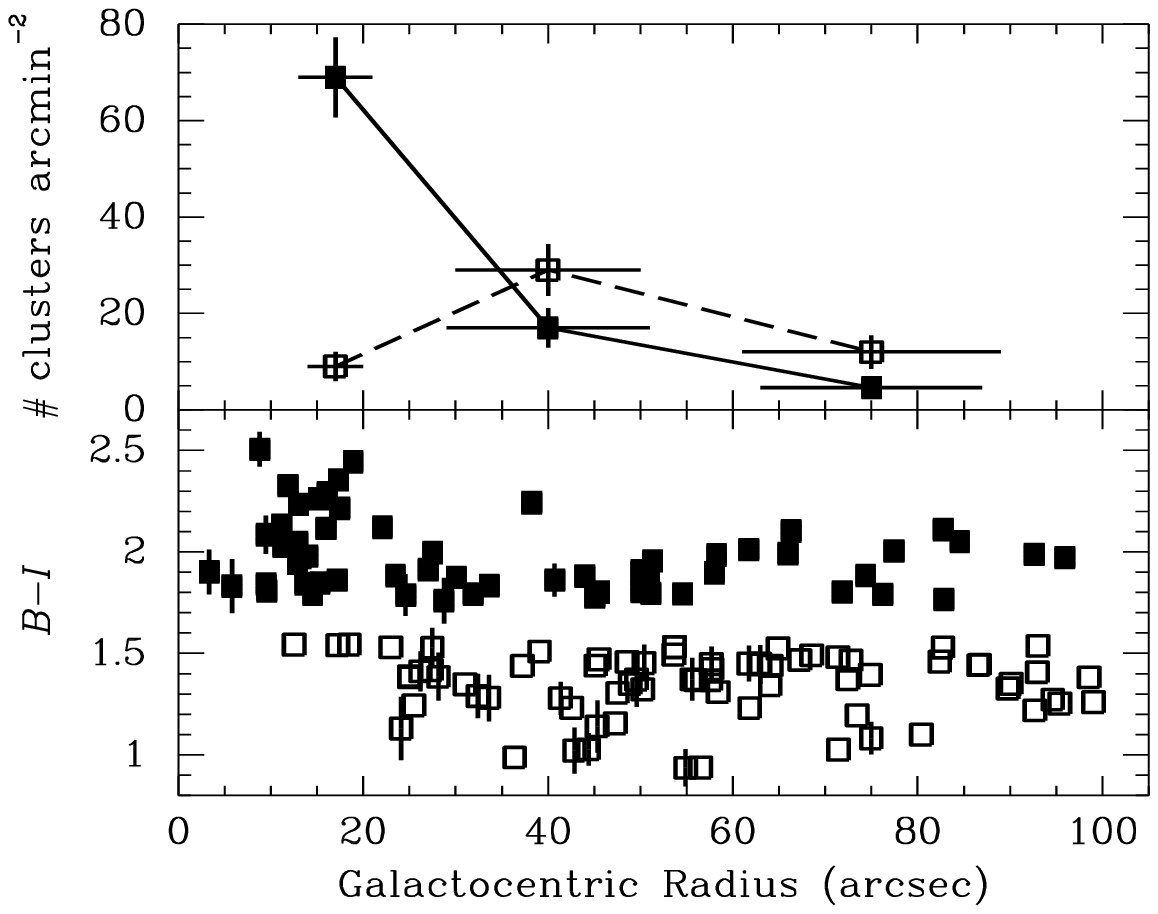,width=8.4cm,angle=-90.}}
\caption[]{Radial distribution of the two cluster subpopulations from the
 WFPC2 data. The lower panel shows \BI\ as a function of galactocentric
 radius, while the upper panel shows surface density as a function of
 galactocentric radius. The surface densities were measured by averaging over
 three logarithmically spaced annuli. 
 Filled squares represent the proposed `metal-rich', second-generation
 clusters and 
 open squares represent the proposed `metal-poor', pre-merger generation of
 clusters. Error bars in the lower panel are only plotted if larger than the
 symbol size. The surface densities have been corrected for foreground stars
 and background galaxies as discussed in Sect.\ \ref{s:S_N}.1.
}
\label{f:b_r_rhoplot}
\end{figure}

\subsubsection{Evolution of \BI\ colour distribution}

Will the colour distribution of the NGC~1316 cluster system evolve into a 
bimodal one, as typically found in giant ellipticals of similar luminosity? 
This is a key question in the context of finding evidence for or against the
`merger scenario' (Ashman \& Zepf 1992) on the nature of bimodal colour
distributions of GC systems of giant elliptical galaxies. 
NGC~1316 constitutes an important probe in this context as well, since {\it
(i)\/} the age and the metallicity of the brightest second-generation GCs are
known to relatively high accuracy (Paper I), and {\it (ii)\/} the first signs
of colour bimodality have already become apparent, with properties that are
entirely consistent with the predictions of the merger scenario. 

So what will the colour distribution look like at a time when the
signs of merger activity have faded away, e.g., after $\sim$\,10 Gyr? This
issue is addressed by panel (b) of Fig.~\ref{f:twopops} which depicts the
time evolution of the absolute magnitude $M_B$ (as well as $B$ at the
distance of NGC~1316) for a globular cluster with the mass of $\omega$\,Cen
($\cal{M}$ = $\mbox{5} \times \mbox{10}^6$ \Mzon, cf. Meylan et al.\
1995). One can see that at solar metallicity and
an age of 3 Gyr, such a cluster is predicted to have $B$ = 21.80, whereas it
will fade to $B$ = 23.05 at an age of 10 Gyr, a fading of 1.25 mag. In that
time span, its \BI\ colour will have reddened from $\sim$\,1.85 to
$\sim$\,2.15 [cf.\ panel (a)]. The peaks in the \BI\ colour
distribution at an age of 10 Gyr are thus expected to be at $B-I$ = 1.55 and
$B-I$ = 2.15. {\it This is remarkably similar to the peaks observed in
bimodal distributions of cluster colours in well-studied giant ellipticals
with luminosities similar to that of NGC~1316\/} (e.g., NGC~1399 and NGC~1404
(Forbes et al.\ 1998), NGC 3923 (Zepf, Ashman \& Geisler 1995b), NGC 4472
(Geisler, Lee \& Kim 1996), NGC 4486 (Kundu et al.\ 1999). Hence, the answer
to the question posed as the first sentence of this subsection is 
{\it affirmative\/}. The cluster system of this merger remnant {\it
will\/} evolve into one with a typical bimodal colour distribution (barring
further merger events), providing strong support in favour of the `merger
scenario'. 

\subsubsection{Radial distribution of the two subpopulations}

Another key prediction of the `merger scenario' is that the newly formed,
metal-rich clusters will be more centrally concentrated than the pre-existing
metal-poor clusters. However, the multi-phase collapse (non-merger) scenario
proposed by Forbes et al.\ \shortcite{forb+97} predicts a very similar
distribution of the two subpopulations, so that the occurrence of this
property in `normal' giant ellipticals featuring bimodal colour distributions
does not provide strong evidence in support of either scenario. 
As discussed in the previous subsection however, GC systems of
intermediate-age merger remnants such as NGC~1316 {\it are\/} important
probes in this context. We compare the radial distribution of the two
subpopulations by using the WFPC2 data, assigning clusters with 1.00 $\le
B\!-\!I \le$ 1.55 and $\Delta(B\!-\!I) \le$ 0.1 to represent the `metal-poor'
ones and clusters with 1.75 $\le B\!-\!I \le$ 2.5 and $\Delta(B\!-\!I) \le$
0.1 to represent the second-generation `metal-rich' ones. These criteria
ensure that the \BI\ interval that overlaps with both subpopulations is
excluded from this exercise. The radial distribution of both subpopulations
is depicted in  Fig.~\ref{f:b_r_rhoplot}. A glance at the lower panel (a
diagram of \BI\ vs.\ radius), already shows clearly that the `metal-rich'
clusters are more centrally concentrated than the `metal-poor' ones. 
This conclusion is endorsed by the upper panel which shows a plot of 
cluster surface density vs.\ radius, derived after counting the number of
clusters within three radial annuli, logarithmically spaced in galactocentric
radius. Note that a similar result was found for NGC~3610, a candidate 4 Gyr
old merger remnant featuring a similarly emerging colour bimodality (Whitmore
et al.\ 1997).

\begin{figure*}
\centerline{\psfig{figure=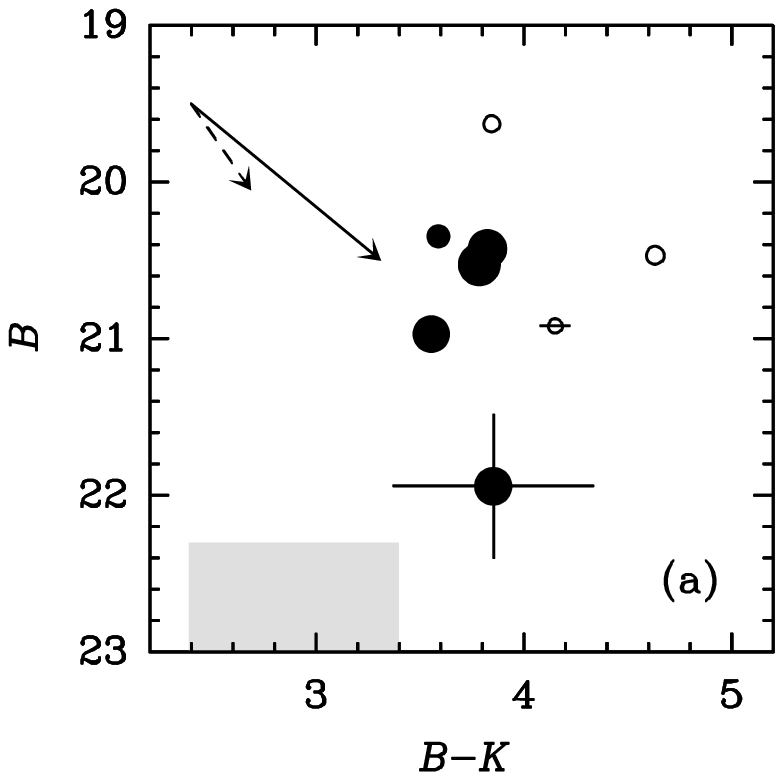,width=6.5cm}
\hspace*{2mm}
\psfig{figure=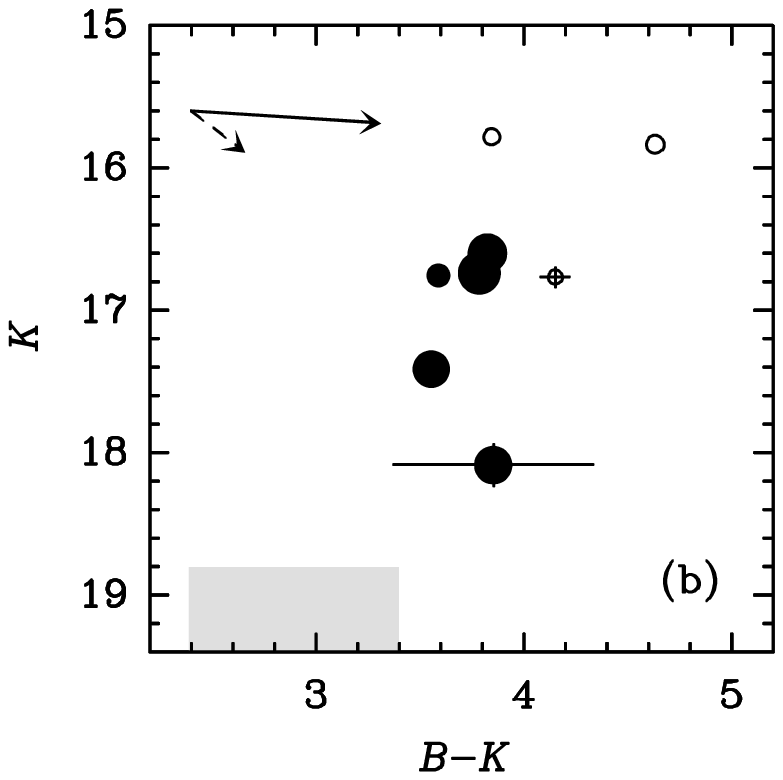,width=6.5cm}}
\caption[]{Optical-IR colour-magnitude diagrams of cluster candidates [$B$
 vs.\ \BK\ in {\bf (a)}; $K$ vs.\ \BK\ in {\bf (b)}]. Symbol types as in
 Fig.~\ref{f:B_BmI_WFPC2}. The symbol {\it sizes\/} have been chosen
 to scale linearly with projected distance to the centre of
 NGC~1316. Error bars are only plotted if they are larger than the
 symbol size. The grey-scale region at the bottom left delineates the
 apparent magnitude and colour range for GCs in our Galaxy, placed at
 the distance of NGC~1316.  
 The solid arrow indicates the reddening vector for $A_B$ = 1 mag of
 extinction, while the dashed arrow indicates the vector for a metallicity
 increase of 0.33 dex at an age of 3 Gyr (see text for details). 
}
\label{f:b_k_bmink}
\end{figure*}

\subsection{Near-IR properties of stellar populations}
\label{s:irpops}

In interpreting the $JHK$ photometry of star cluster candidates in NGC~1316,
we restrict ourselves to the objects that have optical photometry as
well. 
The near-IR photometry was performed in part because it
serves as a very sensitive check for the presence of stellar populations
younger than $\sim$\,2 Gyr, 
which is relevant to the issues at hand in this paper. Near-IR photometry
is remarkably sensitive to the presence of thermally pulsing asymptotic giant
branch (TP-AGB) stars. TP-AGB stars are known to undergo a rather active
phase during which a large amount of fuel in extended hydrogen and helium
shells around degenerate carbon-oxygen cores is burned, which results in a
sharp increase of the near-IR stellar luminosity, particularly in the
$K$-band. This so-called `AGB phase transition' \cite{renbuz86} starts at
ages $\sim$\,200--300 Myr and lasts $\sim$\,1 Gyr (cf.\ Renzini \& Voli 1981;
Iben \& Renzini 1983; 
Maraston 1998; Lan\c{c}on 1999). 

In contrast to near-IR photometry, the evolution of optical colour indices
such as \BV\ or \BI\ for a simple stellar population is monotonic with age,
since they mainly trace the evolution of the turn-off part of the
population. Hence, colour-colour diagrams involving such an 
optical colour index along with an index that is sensitive to the AGB phase
transition (e.g., \VK, \IK, or \JK) can be expected to provide powerful
means of identifying intermediate-age stellar populations. 

However, the actual inclusion of the TP-AGB contribution in population
synthesis models is complicated by poor knowledge of the physical mechanisms 
driving AGB evolution. Specifically, the amount of fuel burned during the
thermally pulsing phase is largely unknown due to uncertainties in the mass
loss rate, mixing, and the efficiency of hydrogen burning at the bottom of
the convective layer (Renzini \& Buzzoni 1986). Fortunately, intermediate-age
star clusters in the Magellanic Clouds cover the relevant age range (see,
e.g., the review by Olszewski 1995) and thus offer an opportunity to
calibrate these effects. This was done by Maraston \shortcite{mara98} who
presented SSP models in which the contribution of TP-AGB stars to the total
SSP energy as a function of age is calibrated on observed photometry of
star clusters in the Magellanic Clouds. Maraston et al.\ (2001)
computed similar models for various metallicities, adopting the
prescriptions by Renzini \& Voli \shortcite{renvol81} to describe the
influence of metallicity on the TP-AGB energy. These models are
described in the next subsection (see Maraston et al.\ 2001 for more details). 

\subsubsection{SSP modelling of populations including TP-AGB stars}

We used the recent sets of SSP models with metallicities of 0.5 \Zsun, 1.0
\Zsun, and 2.0 \Zsun\ that are described in detail in 
Maraston et al.\ \shortcite{mara+01} and Maraston (2001, in preparation). In
summary, these models were computed with the 
evolutionary synthesis code of Maraston \shortcite{mara98}, which is based
upon the fuel consumption theorem (Renzini \& Buzzoni 1986). The
AGB contribution to the bolometric luminosity as a function of age was
calibrated using the observed AGB contributions in Magellanic Cloud clusters
from Frogel, Mould \& Blanco \shortcite{frog+90}, grouped into age bins
adopting the 
classification of Searle, Wilkinson \& Bagnuolo (1980, hereafter SWB). Model
colours were computed by calibrating the partition of the total TP-AGB 
fuel among stars of spectral type C (Carbon stars) and M. The fractional
number of C stars relative to the total number of AGB stars, $N_C$, is a
function of SSP age and metallicity. The age dependence of $N_C$ was
evaluated from the data for Magellanic Cloud clusters \cite{frog+90}, while
the metallicity dependence of $N_C$ was evaluated using the prescription of
Renzini \& Voli \shortcite{renvol81}.  
Observed colours for C-- and M-type AGB stars were taken from
Cohen et al.\ \shortcite{cohe+81}, Cohen \shortcite{cohe82}, and Westerlund
et al.\ \shortcite{west+91}.  

The success of these new models in describing the photometric properties of 
intermediate-age star clusters was shown in Maraston et al.\ (2001,
especially their Fig.\ 5). The star clusters in NGC~7252, a merger remnant
galaxy of age $\sim$\,300--500 Myr (e.g., Hibbard \& Mihos 1995; Schweizer \&
Seitzer 1998), were found to exhibit very red \VK\ colours for their
\BV, which is exactly what is observed for Magellanic Cloud clusters of SWB
types IV--VI. Maraston et al.\ (2001) showed clearly that this behaviour
cannot be understood without the addition of the TP-AGB contribution, while it
was readily fit by the new models for 0.5 \Zsun.  
 
Given the complex and still poorly determined shape of the
spectrum of TP-AGB stars, intermediate-age SSP colours must be
calibrated against observations in every band. For the Cousins
$I$-band, observations of C-- and M-type TP-AGB stars are not available,
nor are integrated $I$ magnitudes of globular clusters in the Magellanic
Clouds. For the purpose of the present paper, the \VI\ model
colours have therefore been calibrated on observations of intermediate-age 
star clusters in NGC~7252 (from Whitmore et al. 1997; see Maraston et
al.\ 2001, Table 2). The dereddened \VI\ colours of these clusters span a
small range:\ \VI\ = 0.64 $\pm$ 0.04. Their ages have been found to be $\sim$
300 Myr (Maraston et al.\ 2001, consistent with the results by Schweizer \&
Seitzer 1998 based on optical spectroscopy). Therefore the model \VI\ at age
300 Myr has been calibrated to reproduce \VI\ = 0.64. Note that this
procedure retains the amount of fuel that was calibrated in Maraston
(1998). The calibrating variable has been chosen to be the \VI\ colour of the
input TP-AGB stars. Specifically, \VI\ = 1.1 for TP-AGB M stars and 1.4 $\la$
\VI\ $\la$ 2.0 for C-type TP-AGB stars is found to perfectly reproduce an
integrated colour \VI\ = 0.64.  

These same input colours for TP-AGB stars have been used to
compute the models older than 300 Myr. This is a reasonable assumption
because the colours of C stars are not found to change significantly
with the age of the parent star cluster (Frogel et al.\ 
1990 and references therein), the evolving parameter being the fuel
consumption.

\subsubsection{Results}

Fig.\ \ref{f:b_k_bmink} shows the $B$ vs.\ \BK\ and $K$ vs.\ \BK\ CMDs
for the detected cluster candidates in NGC~1316. Their location in these
diagrams is compared to the area covered by the system of halo GCs in our
Galaxy (shown in grey scale). The $JHK$ photometry of the Galactic GCs was
taken from Aaronson et al.\ (1978, and private communication). Only data with
photometric errors in \BK\ smaller than 1 mag are plotted. The detected star
clusters are more luminous (as well as redder) than the halo GCs in our
Galaxy, and are thus likely belonging to the second-generation cluster system
formed during the merger. Comparing the two CMDs in Fig.~\ref{f:b_k_bmink}, 
one can see that the colour of the reddest inner clusters (the open circles;
these clusters are located within $\sim$\,20$''$ from the nucleus) is most
likely due to extinction within the dusty filaments rather than being due to
higher metallicity. 
The present near-IR imaging does not reach deep enough to detect the clusters 
belonging to the `blue peak' in the \BI\ histogram (cf.\ Figs.\
\ref{f:BIhists} and \ref{f:twopops}). 8-m class telescopes are needed for
this purpose, which is important to obtain highly increased power (with
respect to optical colours) in disentangling different cluster populations in
galaxies. 

\begin{figure*}
\centerline{
\psfig{figure=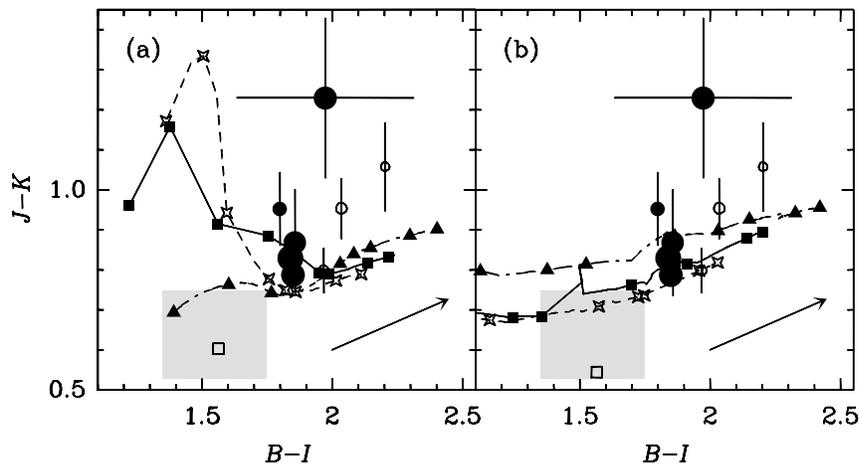,height=6.cm}
}
 \caption[]{\BI\ vs.\ \JK\ colour-colour diagram for star cluster
 candidates in NGC~1316. Symbol types and sizes of the cluster data points as
 in  Fig.~\ref{f:b_k_bmink}.  
 The grey-scale region at the bottom left of each panel delineates the
 observed colour ranges for GCs in our Galaxy. 
 The solid arrow indicates the reddening vector for $A_B$ = 1 mag of
 dust extinction. The left panel {\bf (a)} shows the data along with curves
 and symbols depicting the Maraston (2001) models for a Salpeter (1955)
 IMF, for the metallicities 0.5 solar (dashed line and open triangles), 1.0
 solar (solid line and filled squares) and 2.0 solar (dot-long-dashed line
 and filled triangles). Symbols along the model curves are plotted for ages
 of (from right to left) 14, 10, 4, 3, 2, 1, 0.8, and 0.6 Gyr. 
 The Maraston model for 0.02 solar metallicity and 15 Gyr age is
 shown by an open square for comparison. 
 The right panel {\bf (b)} shows the same data along with 
 curves depicting the BC96 models for a Salpeter IMF. 
 Model metallicities, line types, and symbols are the same as for panel {\bf
 (a)}.  
 The BC96 model for 0.02 solar metallicity and 15 Gyr age is shown by an open
 square for comparison.  
 For the BC96 models, some of the
 symbols corresponding to the youngest ages fall off the plot.  
}
\label{f:bmini_jmink}
\end{figure*}

Fig.~\ref{f:bmini_jmink} shows the \BI\ vs.\ \JK\ colour-colour
diagram of the cluster candidates in NGC~1316. This diagram is particularly
sensitive to the presence of TP-AGB stars, as can be evaluated from
the models of Maraston (2001, described above) which are overplotted in the
left panel for metallicities of 0.5 \Zsun, 1.0 \Zsun, and 2.0 \Zsun. Symbols
along the model curves indicate a number of reference ages. For
comparison, the right panel shows the same data
along with curves describing the BC96 models (which do not include the
calibrated effect of the AGB phase transition). Fig.~\ref{f:bmini_jmink} shows
clearly that the colours of the best-observed clusters in NGC~1316 are indeed
not well described by an intermediate-age population dominated by TP-AGB
stars such as the clusters of NGC~7252. Instead, the NGC~1316 clusters are
best described by the solar metallicity model at an age of 2.5--3 Gyr
(allowing for reddening of the reddest inner clusters by dust as seen
before), again consistent with the spectroscopic results of Paper~I and the
optical photometry discussed in Section~\ref{s:optpops}.  

\subsection{Star cluster luminosity function}
\label{s:LF}

The typical luminosity function (LF) of `old' globular clusters around
galaxies such as our Milky Way or `normal' giant elliptical galaxies such as
NGC~1399 or M\,87 is well represented by a log-normal function (i.e.,
a Gaussian in magnitude units; see e.g., Ashman \& Zepf 1998 and references
therein) of which the peak absolute magnitude concides with that of the
GC system of our Galaxy (e.g., Whitmore 1997). In contrast, the shape of
luminosity functions of clusters in 
`young' [age $\la$ 500 Myr] merger remnant or starburst galaxies is a power
law with a slope of $\sim$\,$-2$ (see, e.g., Meurer et al.\ 1995; Schweizer et
al.\ 1996; Miller et al.\ 1997; Whitmore et al.\ 1999b; Zepf et al.\
1999). This decidedly different behaviour has been used to argue that the
nature of young clusters in mergers or starburst galaxies is fundamentally
different from clusters in old systems \cite{vdb95}. On the other hand
however, several theoretical investigations have shown that dynamical
evolution can significantly alter the mass function of star clusters over time
(e.g., Fall \& Rees 1977; Aguilar, Hut \& Ostriker 1988; Meylan \& Heggie
1997; Murali \& Weinberg 1997; Gnedin \& Ostriker 1997). In fact, dynamical
evolution has been crucial in shaping the LFs of old cluster systems as they
are today (cf.\ Gnedin \& Ostriker 1997). In 
order to study the evolution of cluster LFs by observation as well as theory,
it is important to find systems with ages that are intermediate between those
of the young mergers and the $\ga 10$ Gyr old systems. NGC~1316 is an
interesting probe in this respect, as the bright GCs in NGC~1316 are $\sim$\,3
Gyr old (Paper I).   
For instance, if the GC system of NGC~1316 indeed is a combination of ``old''
GCs that belonged to the merger progenitor galaxies and second-generation GCs
that were formed in the merger, one would expect to find different LFs for
the two subpopulations.

\begin{figure}
\centerline{\psfig{figure=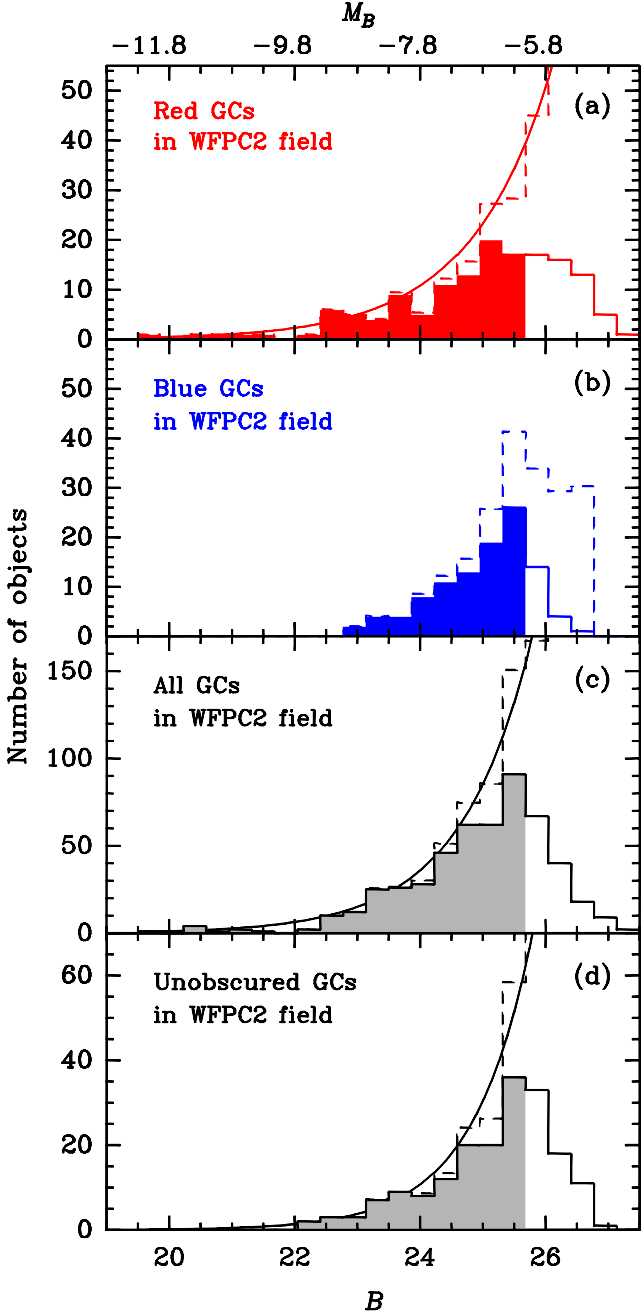,width=8.4cm}}
\caption[]{$B$-band luminosity functions of star cluster candidates using
{\it HST\/} data. The top panel {\bf (a)} shows the luminosity function for
the {\it red\/} cluster candidates on the WFPC2 frames, panel {\bf (b)}
does so for the {\it blue\/} cluster candidates (see discussion in
Section~\ref{s:LF} on the definition of `red' and `blue' clusters in this
context), panel {\bf (c)} does so for {\it all\/} cluster candidates together,
and the bottom panel {\bf (d)}  does so for the star clusters found {\it
outside the dusty central filaments}. 
Bottom axis unit is apparent magnitude $B$, and top axis unit is absolute 
magnitude $M_B$ at the distance of NGC~1316. The histogram is filled for
magnitudes below the 50 per cent completeness limit, and open beyond
it. Dashed lines show the effect of dividing by the completeness fractions
and subtracting the (completeness-corrected) luminosity function of the
background field (cf.\ Elson et al.\ 1998). Solid smooth curves are 
power-law fits to the completeness-corrected luminosity functions.}
\label{f:GCLFs}
\end{figure}

Given the magnitude depth of the {\it HST}/WFPC2 data, its superb power in
discarding background galaxies as star cluster candidates, and the
availability of a background field, we use the {\it HST\/} data for examining
the star cluster LF for NGC~1316. While this item was already discussed to
some extent by Grillmair et al.\ \shortcite{gril+99}, we are interested in
the exact functional representation of the shape of the LF for reasons
outlined above. The LF for the cluster system of NGC~1316 is plotted in Fig.\ 
\ref{f:GCLFs}. 
Four panels are shown: One for all `red' GCs, one for all `blue' GCs (where
`red' and `blue' are defined as for Fig.\ \ref{f:b_r_rhoplot} in Sect.\
\ref{s:optpops}.3), 
one for the {\it full\/} list of cluster candidates found on all WFPC2 chips,
and one for cluster candidates located at radii $> 50''$ from the centre
(i.e., outside the dusty central filaments; hereafter called the `unobscured
sample'). The solid histograms depict the observed LFs, while the dashed
histograms depict the completeness-corrected and background-subtracted LFs. 
The completeness-corrected LF of the full sample of GCs in NGC~1316 can be
seen to continue to increase out to beyond the 50 per cent completeness limit
of the WFPC2 data (cf.\ also Grillmair et al.\ 1999), reminiscent of the LFs
found in young merger remnant galaxies as mentioned above. However, the two
upper panels in Fig.\ \ref{f:GCLFs} clearly show that this behaviour is mostly
due to the `red' GCs, as the LF of the `blue' GCs shows strong evidence for a
`turnover', similar to log-normal LFs such as those found for GCs in our
Galaxy and in `normal' ellipticals.  
For reference, the absolute magnitude of the `turnover' of the Galactic GC
system is $M_B = -$6.6 \cite{harr96}, similar 
to the peak in the LF of the `blue' GCs in NGC~1316. This is consistent with
the notion that the `blue' GCs in NGC~1316 stem from the merger progenitor
spirals. Given the fact that the turnover is near the 50 per cent completeness
level however, this important result should be confirmed with deeper
photometry. 

Given the well-known power-law nature of LFs of young and mixed-age cluster
systems (see, e.g., Elson \& Fall 1985 for the LMC; Whitmore et al.\ 1999b for
NGC~4038/4039), we fitted the completeness-corrected, background-subtracted
LFs for NGC~1316 by a power law of the form   
\[ \phi(L)\,dL \propto L^{\alpha}\,dL\mbox{.} \] 
To limit the impact of uncertainties in the incompleteness factors, the fits
were limited to bins containing objects with $B \le 25.6$. Fig.\ 
\ref{f:GCLFs} shows the results of these fits (solid curves) performed by
least squares with weights proportional to $N$ (the number of objects per
bin). 
The best fit for the `red' GC candidate sample yields $\alpha = -1.23 \pm
0.26$. 
For the `full' GC candidate the fit yields $\alpha = -0.97 \pm 0.09$, while
that for the unobscured sample yields $\alpha = -1.30 \pm 0.16$.  
These LF power-law slopes are significantly flatter than those found in
`young'  mergers (cf.\ $\alpha = -2.6\pm0.2$ for NGC~4038/4039
\cite{whit+99b}, $\alpha = -2.1\pm0.3$ for NGC~3921 \cite{schw+96}, $\alpha =
-1.8\pm0.1$ for NGC~7252 \cite{mill+97}, and $\alpha \sim -1.8$ for NGC~3256
\cite{zepf+99}) and starburst galaxies ($\alpha \sim -2.0$, cf.\ Meurer et
al.\ 1995).  

\subsubsection{Dynamical evolution of star cluster mass function}

The above results indicate that the LF (and mass function) of a star cluster
system of a $\sim$\,3 Gyr old merger remnant such as NGC~1316 is a power law,
with a slope that is significantly flatter than star cluster systems of young
merger remnants. We suggest that this flattening of the power-law LF may
represent an intermediate stage in the dynamical evolution of star cluster
mass functions. We illustrate this below using simple scaling relations
between time scales of known long-term cluster disruption processes. 

The two dynamical processes that are most relevant in determining cluster mass
functions are {\it (i)\/} (internal) evaporation through escaping stars and
{\it (ii)\/} tidal shocking. The characteristic time scales for these
destructive  processes are:   
\begin{equation}
\tau_{\rm evap} \: = \: 0.138 \: \frac{{\cal{M}}_c^{1/2} \,
 r_h^{3/2}}{G^{1/2}\,m_*\,\ln(\Lambda)}
\label{eq:evap}
\end{equation}
for evaporation (Spitzer \& Hart 1971; Gnedin \& Ostriker 1997), and  
\begin{equation}
\tau_{\rm shock} \: = \: \frac{3P_r}{20} \left[ \frac{R_{\rm
 peri}^2\,v_c}{{\cal{M}}_{\rm G}\,\chi(R_{\rm peri})} \right]^2  
 \frac{{\cal{M}}_c}{G \, r_h \left<r_h^2\right> \eta^*} 
 \, \frac{\xi}{\lambda(e, R_{\rm peri})} 
\label{eq:shock}
\end{equation}
for tidal shocking (Aguilar et al.\ 1988). In Eqs.\ (1) and (2), ${\cal{M}}_c$
is the total cluster mass, $r_h$ is the cluster half-mass radius, $G$ is the
gravitational constant, $m_*$ is the average stellar mass in the cluster,
ln($\Lambda$) = ln (0.4$N$) is the Coulomb logarithm ($N$ being the number of
stars in the cluster), $P_r$ is the radial period of the cluster orbit,
$R_{\rm peri}$ is the perigalacticon distance, $v_c$ is the cluster velocity
at perigalacticon, ${\cal{M}}_{\rm G}$ is the galaxy mass, $\chi(R_{\rm
peri})$ is a dimensionless factor that takes 
the mass distribution of the galaxy into account (cf.\ Aguilar \& White
1985), $\left<r_h^2\right>$ is the mean square cluster radius, $\eta^*$ is the
net shock efficiency, $\xi$ is the ratio of the fractional change of cluster 
binding energy to cluster mass induced by the shock (Spitzer \& Chevalier
1972), and $\lambda(e, R_{\rm peri})$ is a dimensionless correction factor
proportional to the square of the difference in force at perigalacticon vs.\ 
at apogalacticon (cf.\ Eq.\ (11) in Aguilar et al.\ 1988). 
For the purpose of this comparative exercise, we neglect the dependence of
$\tau_{\rm shock}$ on the cluster orbit and galaxy potential, i.e., we assume
that we are comparing cluster systems in (giant) ellipticals
covering a relatively small range of galaxy masses (and cluster orbits). This
leaves us with $\tau_{\rm evap} \propto {\cal{M}}_c^{1/2}\,r_h^{3/2}$ and
$\tau_{\rm shock} \propto {\cal{M}}_c \; r_h^{-3}$. 

Imaging studies of young clusters in merger remnants and starburst galaxies
using {\it HST\/} have revealed a very weak dependence of cluster
radii on luminosity (e.g., $r_h \propto L^{\sim 0.07}$ in NGC 3256, Zepf
et al.\ 1999; cf.\ also Meurer et al.\ 1995). Using Zepf et al.'s results, the
relative scalings of destruction time scales as a function of cluster mass
become $\tau_{\rm evap} \propto {\cal{M}}_c^{0.6}$ and $\tau_{\rm shock}
\propto {\cal{M}}_c^{0.8}$. These relations can be used to determine the time
evolution of any characteristic cluster mass that is undergoing a particular
stage of destruction like e.g., the turnover mass of `old' cluster
systems in galaxies such as the Milky Way or nearby `normal' elliptical
galaxies. Assuming an age of 14 Gyr for the old cluster systems, the turnover
mass at an age of 3 Gyr (i.e., the age of the second-generation clusters in
NGC~1316) will be 8 per cent of that in the Milky Way if evaporation is the
dominant destruction mechanism, or 15 per cent of that in the Milky Way if
tidal shocking is the dominant destruction mechanism.  
A relevant question to ask in the context of the observed LF of clusters in
NGC~1316 is then: Would one be able to detect clusters with masses of 
8\,--\,15 per cent of the turnover mass of the Milky Way GC system in the
WFPC2 data? 
According to recent stellar population models and a Salpeter (1955) 
IMF, the ${\cal{M}}/L_B$ ratio of a 3 Gyr old stellar
population of solar metallicity is a factor 1.61 lower than that of
a 14 Gyr old population with 0.05 solar metallicity (Maraston 1999; the
factor is 1.44 when using the BC96 models)\footnote{The dependence on
IMF is negligible in this context: E.g., For a Scalo (1986) IMF, the factor
is 1.57 rather than 1.61.}.  
Given the turnover magnitude of $M_B = -$6.6 in the Milky Way, one would
then expect to see the turnover for the `red' GCs in NGC~1316 occur within the
range $-$5.1~$\la M_B 
\la$~$-$4.4, where the precise absolute magnitude of the turnover mainly
depends on the relative importance of evaporation vs.\ tidal shocking. Note
that this is beyond the 50 per cent completeness limit of the existing 
WFPC2 photometry (cf.\ Fig.\ \ref{f:GCLFs}). Further observations of NGC~1316
are necessary to test whether the turnover in the cluster 
LF indeed appears at the expected luminosity. Resolving this issue should be 
quite easy with the Advanced Camera for Surveys (ACS) which is
currently scheduled to be commissioned in 2002 aboard {\it HST}. 

\subsection{Radial distribution and specific frequency}
\label{s:S_N}

The specific frequency  of GCs, 
\[ S_N \equiv N_{\rm GC} \, 10^{0.4\,(M_V + 15)} \]
\cite{harvdb81}, i.e., the number of star clusters per galaxy luminosity in
units of $M_V$ = $-$15, is an important parameter of star cluster systems of
galaxies. $S_N$ is known to increase systematically along the Hubble sequence,
from $\left< S_N \right> = $0.5 $\pm$ 0.2 for Sc spirals to 2.6 $\pm$ 0.5 for
ellipticals outside galaxy clusters, albeit with considerable galaxy-to-galaxy
scatter \cite{harr91}. As mentioned in Sect.~\ref{s:intro}, a sizable (and
still steadily growing) number of {\it HST\/} observations of merger and
starburst galaxies have led to a wealth of discoveries of young star clusters,
consistent with the predictions of the `merger scenario' for giant ellipticals
(Schweizer 1987; Ashman \& Zepf 1998 and references therein). An interesting
question is then:\ Can the difference in specific frequency between
ellipticals and spirals be accounted for by secondary populations of clusters
created during (gas-rich) mergers? Again, NGC~1316 is a well-suited probe to
test this possibility, in the sense that we know that its body has already
settled down to a typical $r^{1/4}$ surface brightness law, we know the age of
the major merger during which the secondary population of clusters were
created, and it occurred long ago for dynamical evolution of the cluster
system having already had a significant impact, making estimates for its
future evolution less uncertain than for younger mergers. In the remainder of
this section, we derive the current value of $S_N$ for NGC~1316 
and place limits on its future evolution. 

\subsubsection{Radial surface density distribution}

\begin{figure}
\centerline{\psfig{figure=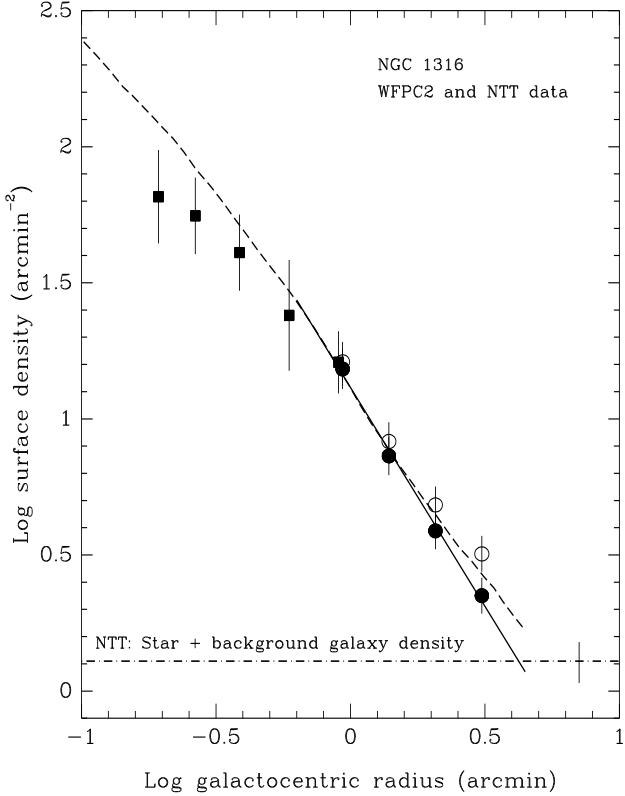,width=8.2cm}} 
\caption[]{Radial surface density profiles for the star cluster system of
 NGC~1316 down to $B$ = 24.0. Filled squares represent measurements from the
 WFPC2 data, corrected for foreground stars and background galaxies as
 explained in the text. Circles represent measurements from the NTT
 data; the open circles have been corrected for foreground stars and
 background galaxies in the same way as the WFPC2 data, while the filled
 circles have been corrected for the density of background galaxies as
 estimated from the CFRS catalog (see text for details). The dot-dashed line
 depicts this estimated density of background galaxies, and its uncertainty is
 indicated by the errorbar at the bottom right of the figure. A power-law
 fit to the background-corrected data beyond the inner `core region' is
 shown by a solid line. The surface brightness profile of the integrated
 $B$-band light of the galaxy (normalized to the surface density of star
 clusters at a radius of 1 arcmin) is shown by the dashed line.}   
\label{f:rhoplot}
\end{figure}

In order to determine the radial surface density distribution and the total
number of clusters in NGC~1316, we use a hybrid method that combines the depth
of the WFPC2 imaging with the much larger sky area covered by the NTT
imaging. The resulting radial surface density distribution is shown in
Fig.~\ref{f:rhoplot}, and was derived as follows. We use
the NTT imaging results for clusters outside the area covered by the dusty
filaments (beyond a radius of 50$''$), while we use the WFPC2 results inside
that radius. Only clusters brighter than the NTT incompleteness limit were
used ($B <$ 24.0). In both datasets, we calculated the number of clusters 
in four annuli, logarithmically spaced in galactocentric radius. Next, we
divided the number of clusters by the appropriate area coverage to yield a
surface density (taking into account the limited azimuthal sky coverage of
the WFPC2 data). Finally, we have to subtract the surface density of
foreground stars and compact background galaxies. For the WFPC2 data, we used
the detected point sources in the field published by Elson et al.\ (1998; see
also Grillmair et al.\ 1999) who observed a background field in the Fornax
galaxy cluster with WFPC2, using the same filters and exposure times as in the
NGC~1316 WFPC2 observations. 
For the NTT data, we considered the surface density of point sources 
in the WFPC2 background field to be a {\it lower limit\/} to the surface
density of foreground stars and background galaxies in the NTT cluster
candidate list, since ground-based seeing doesn't allow one to discriminate
point sources from distant compact galaxies. 
We estimated the level of background galaxy contamination using the
Canada-France Redshift Survey (CFRS) data (Lilly et al.\ 1995), within the
appropriate region of the $B$, $B\!-\!I$ CMD. Background
galaxies at the bright end of our $B$-magnitude range peak at $B\!-\!I
\sim$~1.0 and become bluer towards fainter magnitudes. The resulting estimate
of the surface density of background galaxies down to $B$ = 24.0 is 1.3 $\pm$
0.6 arcmin$^{-2}$ (cf.\ also Elson et al.\ 1998). The background-corrected
surface densities of star clusters from the NTT data (hereafter referred to
as `case A') are shown as filled circles in Fig.~\ref{f:rhoplot}; open 
circles depict the surface densities that are only corrected for the detected
targets in the WFPC2 background field (hereafter referred to as `case
B'). Since some fraction of background galaxies in the NTT field has already
been removed by means of our {\sc daophot-ii} sharpness and roundness
selection criteria, the true surface densities will lie somewhere in between
the open and filled symbols.   

Fig.~\ref{f:rhoplot} depicts the surface density distribution of the
clusters, as well as that of the integrated light of the galaxy for
comparison. The cluster surface density is seen to flatten off (in log-log
space) towards the centre, relative to the integrated light profile (note
that this excludes the possibility of being due to the presence of dust
features in the inner regions). This is a known feature of GC systems of
`normal, old' giant ellipticals, in which this flattening is 
somewhat 
more pronounced (e.g., Grillmair et al.\ 1994; Forbes et al.\ 1998). This 
feature is most probably due to the accumulative effects of tidal shocking
(which is most effective in the central regions, cf.\ Eq.\ \ref{eq:shock}) and 
dynamical friction during the first few dynamical time-scales after the
assembly of the galaxy (see, e.g., Gnedin \& Ostriker 1997). The somewhat more 
moderate flattening of the surface density of the cluster system of NGC~1316
towards the centre relative to that of `old' giant ellipticals is consistent
with the notion that NGC~1316 is only a few Gyr old. 

Excluding the data points within a galactocentric radius of 40$''$, we made
least-square fits to the surface density data using a power law of the form 
\[ \rho(r) = \rho_0 \, r^{\alpha}\mbox{.} \]
The best fit for the background-corrected data (`case A') is parametrized by
$\rho_0$ = (13.0 $\pm$ 1.1) arcmin$^{-2}$ and $\alpha$ = $-$1.61 $\pm$ 0.08,
and is depicted by a solid line in Fig.~\ref{f:rhoplot}. The corresponding
best-fitted slope for the data that were only corrected for the counts
in the WFPC2 background field (`case B') is $\alpha$ = $-$1.36 $\pm$ 0.10,
whereas the slope for the integrated $B$-band light of the galaxy is $\alpha$
= $-$1.41 $\pm$ 0.02 in the same radial range. We conclude that the radial
surface density of the clusters (down to $B$ = 24.0) is consistent with that
of the underlying galaxy light. 
This situation is somewhat different from the case of GC systems of
`normal' giant ellipticals, 
for which the cluster distribution is typically somewhat more extended than
that of the underlying starlight (e.g., Fleming et al.\ 1995; 
Forbes et al.\ 1998; Kundu et al.\ 1999). This is once again consistent
with the intermediate-age nature of the NGC~1316 cluster system in view of the
dynamical evolution-related issues mentioned above. 

\subsubsection{Present-day specific frequency}

Using the radial surface density profile, we can estimate the total number of
clusters $N_{\rm GC}$ and the specific frequency $S_N$. 
In view of the observed radial extent of the
surface density distribution, we decided to use 10 $\pm$ 1 arcmin as the outer
radius for the cluster system (the uncertainty in the radial extension of the
cluster system was chosen somewhat arbitrarily, but in accordance with
numerous previous studies of cluster systems). 
Given the power-law nature of the LF of the cluster system of NGC~1316, the
derivation of $S_N$ is not as straightforward 
as for systems featuring log-normal LFs. In particular, the derived
$S_N$ is a lower limit, as it depends on the applied faint-magnitude limit. 
Integrating the power-law fit to the surface density between log (radius)
= $-$0.2 and +1.0 $\pm$ 0.1 (with radius in units of arcmin), integrating over
the completeness-corrected LF (described above in Section~\ref{s:LF}), and
adding the number of clusters found in the WFPC2 data interior to log (radius)
= $-$0.2 yields the total number of clusters. Table~\ref{t:S_N} lists the
total number of clusters down to $B$ = 25.5 and $B$ = 26.0. 
(beyond which the completeness corrections become too uncertain) both for
cases A and B described above. 
The corresponding specific frequencies are also listed, assuming $V_T^0$ =
8.53 (RC3), equivalent to $M_V$ = $-$23.27.   

\begin{table}
\caption[]{Number $N_{\rm GC}$ and specific frequency $S_N$ of clusters
 in NGC~1316.} 
\label{t:S_N}
\begin{tabular*}{8.4cm}{@{\extracolsep{\fill}}c|cc|cc@{}} \hline \hline
~ & & & & \\ [-1.8ex]
Magnitude limit & $N_{\rm GC}$ & $S_N$ & $N_{\rm GC}$ & $S_N$ \\ [0.2ex]
 & \multicolumn{2}{c|}{(Case A)\rlap{$^*$}} & 
	\multicolumn{2}{c}{(Case B)\rlap{$^*$}} 
 \\ [0.5ex] \hline 
~ & & & & \\ [-1.8ex]
$B \le 25.5$ & 1910 $\pm$ 120 & 1.2 $\pm$ 0.2 & 2805 $\pm$ 225 & 1.5 $\pm$ 0.2
 \\ 
$B \le 26.0$ & 2405 $\pm$ 210 & 1.5 $\pm$ 0.3 & 3535 $\pm$ 400 & 1.9 $\pm$ 0.3
 \\ [0.8ex] \hline
\multicolumn{5}{c}{\ } \\ [-1.8ex]
\end{tabular*}
\smallskip
\parbox{8.4cm}{
\baselineskip=0.98\normalbaselineskip
{\small
\noindent 
{\sl Note to Table \ref{t:S_N}}: $^*$\,Case A represents the case in which
estimated contamination from both foreground stars and background galaxies
(the latter estimate uses data from the CFRS survey) has been corrected for,
while Case B is only corrected for the estimated contamination from foreground
stars. Reality is likely to lie in between the two cases, as discussed in
Sect.~\ref{s:S_N}.}  
}
\end{table}

As discussed above, the `real' specific frequency is most likely
to lie in between the results for cases A and B. Hence, $S_N$ = 1.7 $\pm$ 0.4
is our best estimate for the {\it present-day\/} specific frequency of
clusters with $B \le$~26.0. While this value of $S_N$ is on the low side for
giant ellipticals, it is by no means extraordinary low:\ e.g., 6 giant
ellipticals in the compilation of Harris (1991) have values lower than 1.7. 

\subsubsection{Evolution of the specific frequency}

\begin{figure}
\centerline{\psfig{figure=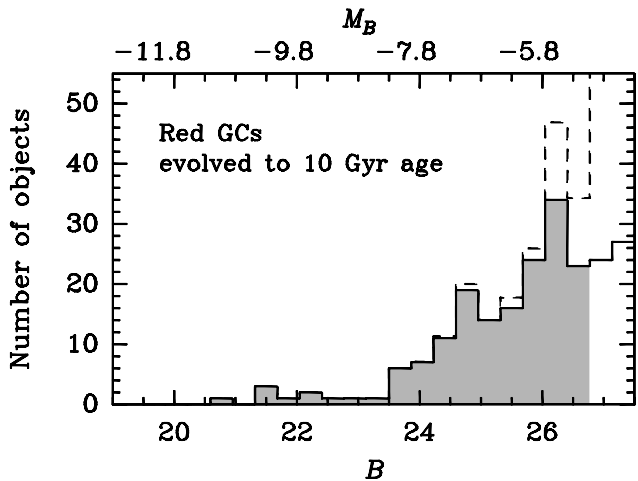,width=8.3cm}}
\caption[]{$B$-band luminosity function of the red cluster candidates in
NGC~1316 after luminosity dimming to an age of 10 Gyr. Hatching and
line types as in Fig.\ \ref{f:GCLFs}.
}
\label{f:GCLF_evolved}
\end{figure}

In order to draw a parallel between the $S_N$ in NGC~1316 and that of
`normal, old' giant ellipticals (e.g., with a luminosity-weighted population
age of $\sim$\,10 Gyr), one has to take into account three main evolutionary
effects:  
{\it (i)\/} 
Luminosity fading of the integrated $V$-band light of NGC~1316,
{\it (ii)\/} 
luminosity fading of the 3-Gyr-old, second-generation clusters, and 
{\it (iii)\/} 
disruption of low-mass clusters in the potential well of NGC~1316.  
As to effect {\it (i)\/} above, recent studies have shown that the
(luminosity-weighted) age of the integrated light of NGC~1316 within
$\sim$\,20$''$ radius is $\sim$\,2 Gyr, as derived from optical line-strength
indices (Kuntschner 2000). 
The age of the integrated light outside 20$''$ radius is not as well
constrained. However, the optical and near-IR colours in those regions are
similar to those of `normal' giant ellipticals (Schweizer 1980; Caon et al.\
1994; Glass 1984), which may indicate that the stellar population there is
similarly `old'. Table~\ref{t:evol_S_N} lists the evolution of $M_V$
with time [using 
the Maraston (1998) SSP models for solar metallicity and a Salpeter IMF; the
BC96 models give very similar results] for two simplified situations:  
(a) assuming that the current age of the integrated light of the
whole galaxy is 2 Gyr, and (b) assuming that the age of the
integrated light in the inner 20$''$ is 2 Gyr, and 14 Gyr outside that. We
consider reality to be most likely nearer the latter assumption. 
As to effect {\it (ii)\/} above, we considered the following. We divided the
cluster 
candidate sample up into `blue' and a `red' subsamples. In this context, the
`blue' subsample was defined as the candidates with 1.0~$\leq$~\BI~$<$~1.65,
whereas the `red' subsample was defined by 1.65~$\leq$~\BI~$<$~2.5. The $B$
magnitudes of the `red' clusters were then faded to an age of 10 Gyr according
to the Maraston (2001) SSP models for solar metallicity and a Salpeter IMF. 
A new LF was created for this evolved `red' GC population, which is shown in
Fig.\ \ref{f:GCLF_evolved}. To estimate the expected turnover magnitude for
this GC population, we considered that for a GC of a given mass, the $B$ 
magnitude for solar metallicity is 1.1 mag fainter than for 0.03 solar (the
median metallicity of the GC system of our Galaxy). Hence, the turnover  
magnitude for a solar metallicity GC system should be at $M_B = -$6.6 + 1.1 = 
$-$5.5 mag at 14 Gyr age, or $M_B = -$5.9 mag at 10 Gyr age (cf.\ Fig.\
\ref{f:twopops}b), corresponding to $B$ = 25.9 at the distance of NGC~1316. 
As to disruption of clusters during the evolution of a GC system, we refer
to the simulations of Vesperini (2000) who showed that the great majority of
the disrupted clusters are those with initial masses $\la$ 10$^{5}$ \Mzon\
(cf.\ Fig.\ 13 of Vesperini 2000), corresponding to $M_B \ga -5.2$ mag for
solar-metallicity GCs at an age of 10 Gyr, which is beyond the
turnover. Hence, this effect should not alter the LF of the clusters that are
brighter than the turnover significantly. 

Considering the above, we estimated the $S_N$ of the GC system of NGC~1316 at
a merger age of 10 Gyr as follows. For the `blue' population, we counted the
GCs up to the absolute magnitude of the turnover of the LF of our Galaxy ($M_B
= -$6.6, i.e., $B$ = 25.2), corrected for incompleteness as a function of
magnitude, and multiplied that number by a factor 2 (i.e., 
assuming a log-normal LF for the evolved GC system). The result is a total of
1950 $\pm$ 140 `blue' clusters. For the `red' population, we counted the
clusters having a `faded' $B$ magnitude up to 25.9 (as discussed above), and
multiplied that number by a factor 2 as well, leading to a total of 1450 $\pm$
145 clusters. The corresponding $S_N$ values are 
listed in Table~\ref{t:evol_S_N}, which 
shows that even if the currently intermediate-age
stellar population in NGC~1316 only inhabits its central part, the specific
cluster frequency will evolve to a value consistent with those of typical 
ellipticals in the field and poor groups (for which $S_N$ = 2.6 $\pm$ 0.5,
Harris 1991) at an age of $\sim$\,10 Gyr.

\begin{table}
\caption[]{Evolution of luminosity and specific frequency of NGC~1316.}
\label{t:evol_S_N}
\begin{tabular*}{8.4cm}{@{\extracolsep{\fill}}c|cc|cc@{}} \hline \hline
~ & & & & \\ [-1.8ex]
Age & $\Delta M_V$ & $S_N$ & $\Delta M_V$ & $S_N$ \\ [0.2ex]
 & \multicolumn{2}{c|}{\sl Age (N1316) = 2 Gyr\rlap{$^*$}} &
 \multicolumn{2}{c}{\sl Age (N1316) = 2 Gyr + 14 Gyr\rlap{$^*$}}  
 \\ [0.5ex] \hline 
~ & & & & \\ [-1.8ex]
  2 Gyr & +0.00 &  1.7 $\pm$ 0.4 & +0.00 & 1.7 $\pm$ 0.4 \\
 10 Gyr & +1.63 & 8.2 $\pm$ 1.9 & +0.14 & 2.1 $\pm$ 0.5 \\ [0.8ex] \hline
 \multicolumn{4}{c}{\ } \\ [-1.8ex]
\end{tabular*}
\smallskip
\parbox{8.4cm}{
\baselineskip=0.98\normalbaselineskip
{\small
\noindent 
{\sl Note to Table \ref{t:evol_S_N}}: $^*$\,The values listed under the
column header `Age (N1316) = 2 Gyr' were calculated under the
assumption that all integrated light of NGC~1316 is 2 Gyr old, while the
values listed under `Age (N1316) = 2 Gyr + 14 Gyr' were calculated
under the assumption that only the integrated light within a galactocentric
radius of 20$''$ (which is currently 15\% of the total light) is 2 Gyr old,
while the rest is 14 Gyr old.} 
}
\end{table}

\section{Summary}
\label{s:summ}

Using a combination of {\it HST\/} and large-field ground-based imaging of
NGC~1316, a well-known early-type merger remnant galaxy for which we
determined an age of $\sim$\,3 Gyr from spectra of its brightest GCs in Paper
I, we presented the optical and near-IR photometric properties of its star
cluster system in this paper. Our main results are itemized below, followed
by a short discussion. 
\begin{itemize}
\item 
The \BI\ colour distribution of the clusters exhibits two peaks: One at \BI\
$\sim$ 1.5 (coinciding with the peak \BI\ colour of the Galactic GC system)
and one at \BI\ $\sim$ 1.8. The brightest clusters mainly populate the red
peak, while the blue peak gets more populated going towards fainter
magnitudes. The clusters in the red peak make up about 50--60 per cent of the
total cluster population. 
\item
By comparing the photometric properties of the brightest 8 clusters with the
those of the Galactic GC system as well as with current SSP models which
include the contribution of the AGB phase of stellar evolution,
we find that 
 \begin{enumerate}
 \item 
 They are up to an order of magnitude more luminous
 than $\omega$\,Cen, the most massive GC in our Galaxy, and hence are most
 probably significantly younger than 10--14 Gyr; 
 \item 
 Their optical-near-IR colours are {\it not\/} as red as intermediate-age
 (0.3--1.5 Gyr old) GCs such as found in the Magellanic Clouds and in
 NGC~7252 (which are due to the presence of TP-AGB stars). Instead, their
 optical and near-IR colours are all best fit by a 2.5--3 Gyr old population
 of roughly solar metallicity, consistent with the results from our earlier
 spectral analysis (cf.\ Paper I). 
 \end{enumerate}
Together with the optical properties of the cluster system described in the
previous item, these results are consistent with a scenario involving the
presence of two cluster subpopulations in NGC~1316: {\it (i)\/} A `red' (\BI\
$\sim$ 1.8) population of clusters which were created during a major gas-rich
merger that occurred $\sim$ 3 Gyr ago, having solar metallicity to within
$\pm$ 0.15--0.2 dex (as derived from spectra in Paper I), and {\it (ii)\/} a
`blue' (\BI\ $\sim$ 1.5) population of `old',  metal-poor clusters that were
associated with the halos of the pre-merger galaxies.  
We show that under this assumption, the `red' clusters will evolve to
$B\!-\!I \sim\;$2.15 in a time span of 10 Gyr, which colour is remarkably
similar to those of the red peaks observed in bimodal distributions of
cluster colours in well-studied `normal, old' giant ellipticals with galaxy
luminosities similar to that of NGC~1316, suggesting that the
red clusters in giant ellipticals have a similar origin. 
\item 
The luminosity function (LF) of the cluster population is best represented by
a power law with a slope $\alpha = -1.2 \pm 0.3$. The power-law behaviour is
dominated by the `red' clusters; the LF of clusters populating the blue end of
the colour distribution shows evidence for a `turnover' as found in the GC
systems of nearby galaxies like the Milky Way and M\,31. The power-law slope
of the GC system of NGC~1316 
is flatter than those found in `young' mergers with ages $\la$ 500 Myr. We
suggest that this flattening of the power-law LF slope represents a stage in
the dynamical evolution of star cluster mass functions that is intermediate
between those found in `young' merger remnants and those found in `normal',
old giant ellipticals (which show a log-normal LF). Using simple scaling
relations between time-scales of long-term cluster disruption processes (tidal
shocking and evaporation), we estimate the absolute magnitude of the turnover
of the LF of a 3 Gyr old cluster population to be in the range $-$5.1~$\la M_B
\la$~$-$4.4, which is beyond the 50\% completeness limit of the available
WFPC2 data. 
\item 
The radial distribution of the surface number density of NGC~1316's star
cluster system is consistent with that of its integrated light in the outer
regions, while it flattens off slightly towards the centre relative to 
the integrated light. The latter flattening towards the centre is
significantly less pronounced than in `old' giant ellipticals, in which the
outer surface density profile of clusters is typically more extended than
that of the underlying starlight. These findings suggest that the cluster
system of NGC~1316 (and, by analogy, of giant ellipticals in general),
originally experienced the same violent relaxation as did the main body of
the merger remnant, after which tidal shocking of (mosty inner) clusters
cause the cluster surface density profile to become gradually flatter with
time. 
\item 
The present-day specific frequency of clusters with $B \le 26.0$ is $S_N$ =
1.7 $\pm$ 0.4. After taking into account age dimming of the 3-Gyr-old clusters
as well as of the integrated light of the galaxy to an age of 10 Gyr 
[under the conservative assumption that only the inner 20$''$ radius of the
stellar population of NGC~1316 has a luminosity-weighted age of 2 Gyr (as
derived spectroscopically by Kuntschner 2000) and that the rest of the light
is from 14 Gyr old stars], the specific frequency will increase to 2.1 $\pm$
0.5, which is consistent with those found in ellipticals in the field and poor
groups.   
\end{itemize}

All these features of the star cluster system of NGC~1316 constitute
important evidence in support of `merger scenarios' for forming giant
elliptical galaxies through major gas-rich mergers. I.e., 
the system of luminous, metal-rich clusters in NGC~1316 is consistent with
being a recently-formed analog of the `red' clusters in `normal' giant
elliptical galaxies. 

\section{Outstanding issues}
\label{s:outlook}

The presence of very luminous second-generation GCs in NGC~1316 that are of
an age at which dynamical evolution is likely to have already had a
significant impact is relevant to a number of outstanding issues of interest
going beyond the particular case of NGC~1316.  
One such issue is whether the mass function of clusters created in mergers is
biased towards higher masses than that of clusters in `normal' galaxies. 
E.g., W3, the brightest cluster in NGC~7252 (Whitmore et al.\ 1993)
has a reported absolute magnitude $M_V^0 = -16.2$ (Miller et al.\ 1997). At
its age of $\sim$\,300 Myr and its metallicity of 0.5 \Zsun, Maraston et al.\
\shortcite{mara+01} derive a mass of $\sim$\,3.7 $\times$ 10$^7$ \Mzon\ (for
a Salpeter IMF), 
which is $\sim$\,10 times higher than the mass of $\omega$\,Cen. Even for a
significantly older merger remnant such as NGC~1316, the brightest
cluster (designated \#\,114 in Paper I) has $M_V^0 = -13.0$, which translates
to a mass of (1.4 $\pm$ 0.2) $\times$ 10$^7$ \Mzon\ at an age of 3.0 $\pm$
0.5 Gyr and solar metallicity (for a Salpeter IMF):\ still a factor $\sim$\,5
higher than the mass of $\omega$\,Cen. The masses of these bright clusters
will be checked by high-dispersion dynamical measurements using the ESO VLT in
order to enable further study of mass functions of GCs created in
mergers. This will be the subject of a forthcoming paper. 

Another important outstanding issue is whether or not such second-generation
clusters show supersolar [$\alpha$/Fe] element ratios. The stellar
populations of many `normal, old' giant ellipticals are known to be
$\alpha$-enhanced by $\sim$ 0.2--0.3 dex (Worthey, Faber \& Gonz\'alez 1992;
Davies, Sadler \& Peletier 1993; Kuntschner 2000). Recent models of galaxy
formation that incorporate the evolution of chemical enrichment of the ISM
during bursts (Thomas, Greggio \& Bender 1999) show that such
$\alpha$-element enhancements can be reached by merging of small sub-galactic
entities on a short time scale ($\sim$\,1 Gyr) so that much of the chemical
enrichment of the ISM is performed {\it in situ\/} by type II
supernovae, which are the main producers of $\alpha$ elements. 
Hence, if major gas-rich mergers are indeed responsible for forming giant
elliptical galaxies (and the second-generation clusters), those mergers 
most likely occurred up to a few Gyr after the pre-merger
galaxies were created. This allowed time for the latter's ISM to become
sufficiently enriched (most probably primarily by SN\,II ejecta, cf.\ Thomas
et al.\ 1999). New observations should be conducted to find out whether or
not the `red' clusters in `normal' giant ellipticals show supersolar $\alpha$
element abundances. If found, this will constitute an important constraint to
merger models such as the well-known `hierarchical' merger model for
elliptical galaxies (e.g., Kauffmann 1996) in the sense that the building
blocks that form giant elliptical galaxies must have already existed
$\sim$\,10$^9$ yr after the `halo' GCs around galaxies were formed.   

However, merging of two sizable gas-rich spirals at {\it late\/} epochs (which
is relevant to young and intermediate-age merger remnants such as NGC~1316 and
NGC~7252) is {\it not\/} expected to produce supersolar $\alpha$-element
abundances, because most of the stars in the merging spiral disks have formed
in a long-lasting ($\sim$\,10 Gyr) star formation process, leading to
approximately solar abundances and element ratios in their ISM. If stars form
out of such gas with a Salpeter IMF, they are expected to show roughly solar
[$\alpha$/Fe] ratios. In such mergers at late epochs, $\alpha$-enhancement
only occurs in models featuring a significantly flattened IMF during the
major starburst following the merger (Thomas et al.\ 1999). At an age of
$\sim$\,3 Gyr, 
the spectral features of Magnesium and Iron in the blue-visual region should
be easily detectable. Thus, the bright clusters in NGC~1316 
are very interesting probes to test such scenarios for chemical enrichment of
merging galaxies. 

\paragraph*{Acknowledgments.} \ \\ 
This work is based on observations obtained at the European Southern
Observatory, La Silla, Chile (Observing Programme 58.E--0666), and on
archival data of the NASA/ESA {\it Hubble Space Telescope}, which is
operated by AURA, Inc., under NASA contract NAS 5--26555. 
We are grateful to the European South\-ern Observatory for allocating
observing time to this project, and for having done a great job of archiving
observations made with the NTT since its inauguration. We thank Chris Lidman
and the technical support staff of the European Southern Observatory for their
support during the near-IR observations, Brad Whitmore and Markus
Kissler--Patig for very useful discussions, and the referee Jean
Brodie for comments that led to an improvement of the paper. 
MVA was partially supported by grants from CONICET, Agencia Nacional de
Promoci\'on Cient\'{\i}fica y Tecnol\'ogica and Secretar\'{\i}a de Ciencia y
T\'ecnica de la Universidad Nacional de C\'ordoba, Argentina. MVA also thanks
the Chilean Fondecyt for a visitor grant.  
CM is supported by the `Sonderforschungsbereich 375-95 f\"ur
Astro-Teilchenphysik' of the Deutsche Forschungsgemeinschaft. 
DM's work was performed in part under the auspices of the Chilean Fondecyt
Nos.\ 01990440 and 07990048, and of DIPUC No.\ 98.16e.

\appendix

\section{Photometry tables of star cluster candidates}

%
%
\begin{table*}
\small
\caption[]{Photometry and astrometry of the 50 brightest GC candidates 
on the WFPC2 frames. $JHK$ photometry is from 
the 2.2-m ESO/MPI telescope. The object list is sorted on $B$
magnitude (brightest first).} 
\label{t:WFPC2phot}
\begin{tabular*}{14cm}{@{\extracolsep{\fill}}rrccccc@{}} \hline \hline
~ & & & & \\ [-1.8ex]
\multicolumn{1}{c}{$\Delta$ RA} & \multicolumn{1}{c}{$\Delta$ DEC} & 
$B$ & \BI\ & $K$ & \JK\ & \HK\ \\ [0.2ex]
\multicolumn{1}{c}{arcsec} & \multicolumn{1}{c}{arcsec} & 
mag & mag & mag & mag & mag \\ [0.5ex] \hline 
~ & & & & \\ [-1.8ex]
      8.91  & $-$11.46 &   19.627 $\pm$ 0.003  &  1.866  $\pm$  0.005  &  15.78 $\pm$ 0.04 &  0.83  $\pm$ 0.06 &  0.15  $\pm$ 0.06 \\      
  $-$48.09  & $-$22.24 &   20.241 $\pm$ 0.003  &  1.744  $\pm$  0.004  &  16.60 $\pm$ 0.06 &  0.78  $\pm$ 0.08 &  0.11  $\pm$ 0.08 \\
  $-$25.01  & $-$13.04 &   20.343 $\pm$ 0.003  &  1.700  $\pm$  0.005  &  16.75 $\pm$ 0.07 &  0.95  $\pm$ 0.09 &  0.18  $\pm$ 0.09 \\
     12.17  &    12.14 &   20.467 $\pm$ 0.003  &  1.935  $\pm$  0.005  &  15.83 $\pm$ 0.05 &  0.95  $\pm$ 0.08 &  0.35  $\pm$ 0.07 \\
  $-$43.94  & $-$38.32 &   20.523 $\pm$ 0.003  &  1.740  $\pm$  0.005  &  16.73 $\pm$ 0.06 &  0.83  $\pm$ 0.08 &  0.28  $\pm$ 0.08 \\
~ & & & & \\ [-1.8ex]		       			      			      		           		         
      4.13  &    10.50 &   20.918 $\pm$ 0.005  &  2.103  $\pm$  0.005  &  16.74 $\pm$ 0.07 &  1.06  $\pm$ 0.11 &  0.22  $\pm$ 0.10 \\
  $-$45.15  & $-$21.03 &   20.967 $\pm$ 0.005  &  1.757  $\pm$  0.006  &  17.41 $\pm$ 0.10 &  0.86  $\pm$ 0.13 &  0.23  $\pm$ 0.13 \\
     53.03  &    23.58 &   21.313 $\pm$ 0.007  &  1.972  $\pm$  0.008 \\
  $-$12.77  &  $-$3.75 &   21.609 $\pm$ 0.008  &  2.047  $\pm$  0.010 \\
  $-$59.81  & $-$33.74 &   22.094 $\pm$ 0.008  &  1.721  $\pm$  0.011 \\
~ & & & & \\ [-1.8ex]		       			      
      3.99  &    74.26 &   22.373 $\pm$ 0.009  &  1.959  $\pm$  0.011 \\
  $-$15.88  &  $-$9.49 &   22.413 $\pm$ 0.012  &  1.618  $\pm$  0.017 \\
  $-$49.09  & $-$31.98 &   22.430 $\pm$ 0.010  &  1.754  $\pm$  0.013 \\
      1.02  & $-$13.66 &   22.483 $\pm$ 0.009  &  1.918  $\pm$  0.013 \\
     37.20  &    72.27 &   22.654 $\pm$ 0.010  &  1.740  $\pm$  0.013 \\
~ & & & & \\ [-1.8ex]		       			      
      1.74  &    33.59 &   22.656 $\pm$ 0.013  &  1.909  $\pm$  0.015 \\
  $-$63.76  &    21.37 &   22.657 $\pm$ 0.011  &  1.771  $\pm$  0.014	\\
     20.40  &    63.09 &   22.706 $\pm$ 0.012  &  2.178  $\pm$  0.013	\\
  $-$25.66  &  $-$9.95 &   22.729 $\pm$ 0.016  &  2.071  $\pm$  0.019	\\
      2.52  &  $-$9.22 &   22.743 $\pm$ 0.015  &  1.883  $\pm$  0.022	\\
~ & & & & \\ [-1.8ex]		       			      
     43.14  &    72.73 &   22.770 $\pm$ 0.011  &  2.125  $\pm$  0.013 \\
  $-$23.15  & $-$39.81 &   22.870 $\pm$ 0.015  &  1.813  $\pm$  0.019	\\
     13.54  &    18.60 &   22.908 $\pm$ 0.023  &  1.605  $\pm$  0.028	\\
   $-$6.76  & $-$15.93 &   22.950 $\pm$ 0.014  &  2.430  $\pm$  0.016	\\
  $-$38.30  &    64.33 &   22.951 $\pm$ 0.013  &  1.469  $\pm$  0.017	\\
~ & & & & \\ [-1.8ex]		       			      
     29.16  &    46.12 &   23.006 $\pm$ 0.016  &  1.868  $\pm$  0.018 \\
   $-$5.48  &  $-$7.77 &   23.030 $\pm$ 0.021  &  1.916  $\pm$  0.025	\\
  $-$41.55  & $-$23.01 &   23.041 $\pm$ 0.017  &  1.677  $\pm$  0.021	\\
  $-$62.41  & $-$54.32 &   23.066 $\pm$ 0.015  &  1.604  $\pm$  0.018	\\
  $-$24.13  &     2.52 &   23.074 $\pm$ 0.022  &  1.781  $\pm$  0.026	\\
~ & & & & \\ [-1.8ex]		       			      
  $-$24.49  &    78.68 &   23.089 $\pm$ 0.013  &  1.534  $\pm$  0.017 \\
   $-$8.07  &  $-$7.71 &   23.131 $\pm$ 0.022  &  2.207  $\pm$  0.025	\\
  $-$28.74  & $-$13.85 &   23.134 $\pm$ 0.022  &  1.867  $\pm$  0.026	\\
  $-$21.48  & $-$42.39 &   23.197 $\pm$ 0.021  &  1.381  $\pm$  0.027	\\
  $-$27.23  &  $-$2.73 &   23.256 $\pm$ 0.026  &  1.493  $\pm$  0.035	\\
~ & & & & \\ [-1.8ex]		       			      
  $-$52.60  & $-$13.63 &   23.276 $\pm$ 0.020  &  1.636  $\pm$  0.024 \\
     15.78  &  $-$6.93 &   23.278 $\pm$ 0.023  &  1.615  $\pm$  0.030	\\
  $-$34.60  & $-$27.10 &   23.288 $\pm$ 0.023  &  1.754  $\pm$  0.027	\\
  $-$21.84  &    75.37 &   23.292 $\pm$ 0.017  &  1.644  $\pm$  0.021	\\
   $-$0.27  &    13.01 &   23.295 $\pm$ 0.022  &  2.309  $\pm$  0.025	\\
~ & & & & \\ [-1.8ex]		       			      
  $-$36.77  &     5.88 &   23.302 $\pm$ 0.021  &  1.514  $\pm$  0.028 \\
  $-$82.63  & $-$46.10 &   23.302 $\pm$ 0.016  &  1.642  $\pm$  0.020	\\
     13.36  &    19.33 &   23.311 $\pm$ 0.033  &  1.958  $\pm$  0.039	\\
     53.16  &    79.19 &   23.320 $\pm$ 0.018  &  1.328  $\pm$  0.023	\\
  $-$41.75  &    38.95 &   23.327 $\pm$ 0.018  &  1.674  $\pm$  0.023	\\
~ & & & & \\ [-1.8ex]		       			      
  $-$85.71  &    52.72 &   23.350 $\pm$ 0.015  &  1.722  $\pm$  0.019 \\
  $-$21.31  &    13.12 &   23.352 $\pm$ 0.026  &  1.464  $\pm$  0.034	\\
     12.11  &    45.72 &   23.379 $\pm$ 0.019  &  1.230  $\pm$  0.028	\\
  $-$54.35  & $-$52.15 &   23.389 $\pm$ 0.023  &  1.790  $\pm$  0.027	\\
  $-$49.59  &     7.15 &   23.394 $\pm$ 0.019  &  1.641  $\pm$  0.026	
 \\ [0.8ex] \hline
\multicolumn{5}{c}{\ } \\ [-1.8ex]
\end{tabular*}
\smallskip
\parbox{14cm}{
{\small
\noindent 
{\sl Note to Table \ref{t:WFPC2phot}}: The first two columns give the 
positional offsets of the GC candidates from the centre of NGC~1316 
in RA and DEC, respectively.}}
\end{table*}

%
%
\begin{table*}
\small
\caption[]{Photometry and astrometry of the 50 brightest GC candidates 
on the ESO NTT images. GC candidates that were also present on the WFPC2 
images have been excluded from this list. $JHK$ photometry is from 
the 2.2-m ESO/MPI telescope. The object list is sorted on $B$
magnitude (brightest first).} 
\label{t:NTTphot}
\begin{tabular*}{15.5cm}{@{\extracolsep{\fill}}rrcccccc@{}} \hline \hline
~ & & & & \\ [-1.8ex]
\multicolumn{1}{c}{$\Delta$ RA} & \multicolumn{1}{c}{$\Delta$ DEC} & 
$B$ & \BV\ & \BI\ & $K$ & \JK\ & \HK \\ [0.2ex]
\multicolumn{1}{c}{arcsec} & \multicolumn{1}{c}{arcsec} & 
mag & mag & mag & mag & mag & mag \\ [0.5ex] \hline 
~ & & & & \\ [-1.8ex]
 $-$158.09 &     64.95 &  20.17 $\pm$  0.19 &  0.98  $\pm$ 0.20 &  1.78  $\pm$ 0.20 \\
     50.75 &  $-$46.88 &  20.64 $\pm$  0.23 &  0.87  $\pm$ 0.25 &  1.45  $\pm$ 0.25 \\    
  $-$84.07 & $-$163.18 &  20.71 $\pm$  0.24 &  0.97  $\pm$ 0.26 &  1.73  $\pm$ 0.25 \\    
    148.33 &    132.35 &  20.79 $\pm$  0.25 &  0.90  $\pm$ 0.27 &  1.67  $\pm$ 0.26 \\
     53.89 &     23.29 &  20.83 $\pm$  0.01 &  1.04  $\pm$ 0.09 &  1.82  $\pm$ 0.01 \\
~ & & & & \\ [-1.8ex]		      		           		         
     53.89 &     23.29 &  20.83 $\pm$  0.26 &  1.04  $\pm$ 0.27 &  1.82  $\pm$ 0.27 \\
  $-$60.97 &     94.40 &  20.88 $\pm$  0.26 &  0.93  $\pm$ 0.28 &  1.70  $\pm$ 0.28 \\
     67.31 &     54.96 &  21.03 $\pm$  0.28 &  0.78  $\pm$ 0.30 &  1.37  $\pm$ 0.30 \\
     67.31 &     54.96 &  21.03 $\pm$  0.28 &  0.78  $\pm$ 0.30 &  1.37  $\pm$ 0.30 \\
    178.46 &     52.70 &  21.15 $\pm$  0.30 &  0.72  $\pm$ 0.32 &  1.13  $\pm$ 0.32 \\
~ & & & & \\ [-1.8ex]		      		           		         
     86.04 &    191.25 &  21.22 $\pm$  0.31 &  0.72  $\pm$ 0.33 &  1.01  $\pm$ 0.34 \\
    160.04 & $-$150.52 &  21.23 $\pm$  0.31 &  0.64  $\pm$ 0.34 &  0.99  $\pm$ 0.34 \\    
 $-$123.07 &      8.43 &  21.50 $\pm$  0.35 &  1.01  $\pm$ 0.37 &  1.83  $\pm$ 0.36 \\
     40.89 & $-$198.75 &  21.57 $\pm$  0.36 &  0.68  $\pm$ 0.39 &  1.02  $\pm$ 0.39 \\    
 $-$141.71 &   $-$0.40 &  21.60 $\pm$  0.36 &  1.20  $\pm$ 0.38 &  2.06  $\pm$ 0.38 \\
~ & & & & \\ [-1.8ex]		      		           		         
     12.87 &    129.47 &  21.71 $\pm$  0.38 &  0.86  $\pm$ 0.41 &  1.51  $\pm$ 0.41 \\
    203.34 &     21.51 &  21.73 $\pm$  0.39 &  0.72  $\pm$ 0.42 &  0.99  $\pm$ 0.43 \\
     71.17 &     20.14 &  21.75 $\pm$  0.39 &  0.85  $\pm$ 0.42 &  1.64  $\pm$ 0.41 \\
 $-$127.28 &  $-$31.86 &  21.79 $\pm$  0.40 &  0.97  $\pm$ 0.42 &  1.71  $\pm$ 0.42 \\    
     60.05 &     17.62 &  21.83 $\pm$  0.41 &  0.97  $\pm$ 0.44 &  1.71  $\pm$ 0.43 \\
~ & & & & \\ [-1.8ex]		      		           		         
      5.25 &     74.90 &  21.96 $\pm$  0.01 &  0.85  $\pm$ 0.17 &  1.57  $\pm$ 0.01 \\
    129.89 & $-$142.12 &  21.97 $\pm$  0.44 &  0.31  $\pm$ 0.49 &  1.23  $\pm$ 0.47 \\    
     93.83 &     49.74 &  21.97 $\pm$  0.43 &  0.92  $\pm$ 0.46 &  1.71  $\pm$ 0.46 \\
 $-$171.13 & $-$176.59 &  22.00 $\pm$  0.44 &  1.09  $\pm$ 0.46 &  1.99  $\pm$ 0.46 \\    
  $-$15.98 &  $-$49.92 &  22.07 $\pm$  0.45 &  0.99  $\pm$ 0.48 &  1.86  $\pm$ 0.47 \\    
~ & & & & \\ [-1.8ex]		      		           		         
 $-$140.28 &     58.05 &  22.08 $\pm$  0.46 &  0.99  $\pm$ 0.49 &  1.73  $\pm$ 0.48 \\
   $-$9.71 &  $-$49.94 &  22.10 $\pm$  0.46 &  0.96  $\pm$ 0.49 &  1.82  $\pm$ 0.48 & 18.08 $\pm$ 0.15 &  1.23 $\pm$ 0.20 &  0.50 $\pm$ 0.20  \\
     94.24 &      5.95 &  22.15 $\pm$  0.47 &  1.14  $\pm$ 0.50 &  2.02  $\pm$ 0.49 \\
 $-$133.37 & $-$106.62 &  22.16 $\pm$  0.47 &  0.87  $\pm$ 0.51 &  1.52  $\pm$ 0.50 \\	    
 $-$113.18 &     80.14 &  22.17 $\pm$  0.47 &  0.90  $\pm$ 0.51 &  1.62  $\pm$ 0.50 \\
~ & & & & \\ [-1.8ex]		      		           		         
    190.76 &      9.15 &  22.21 $\pm$  0.49 &  0.56  $\pm$ 0.53 &  1.38  $\pm$ 0.52 \\
 $-$119.69 &   $-$7.11 &  22.34 $\pm$  0.51 &  1.02  $\pm$ 0.55 &  1.78  $\pm$ 0.54 \\
  $-$76.75 &  $-$85.49 &  22.50 $\pm$  0.55 &  1.02  $\pm$ 0.59 &  1.59  $\pm$ 0.58 \\	    
     69.04 &   $-$6.28 &  22.51 $\pm$  0.56 &  0.84  $\pm$ 0.60 &  1.58  $\pm$ 0.59 \\
 $-$115.13 &    112.73 &  22.51 $\pm$  0.56 &  1.01  $\pm$ 0.59 &  1.79  $\pm$ 0.58 \\
~ & & & & \\ [-1.8ex]		      		           		         
 $-$120.31 &  $-$70.95 &  22.55 $\pm$  0.57 &  0.91  $\pm$ 0.61 &  1.69  $\pm$ 0.60 \\	    
  $-$93.24 &   $-$2.41 &  22.57 $\pm$  0.57 &  1.01  $\pm$ 0.61 &  1.77  $\pm$ 0.60 \\
 $-$100.24 &     77.63 &  22.59 $\pm$  0.57 &  1.05  $\pm$ 0.61 &  1.91  $\pm$ 0.60 \\
 $-$110.05 & $-$148.12 &  22.60 $\pm$  0.58 &  1.07  $\pm$ 0.61 &  1.82  $\pm$ 0.61 \\	    
 $-$156.04 &  $-$17.66 &  22.60 $\pm$  0.58 &  0.79  $\pm$ 0.63 &  1.34  $\pm$ 0.62 \\
~ & & & & \\ [-1.8ex]		      		           		         
    122.96 &     82.64 &  22.61 $\pm$  0.59 &  0.61  $\pm$ 0.64 &  1.22  $\pm$ 0.63 \\
  $-$33.49 & $-$164.70 &  22.63 $\pm$  0.59 &  0.91  $\pm$ 0.63 &  1.64  $\pm$ 0.62 \\	    
  $-$37.41 &  $-$59.81 &  22.63 $\pm$  0.59 &  0.99  $\pm$ 0.63 &  1.54  $\pm$ 0.63 \\	    
    110.55 &      1.67 &  22.67 $\pm$  0.60 &  1.08  $\pm$ 0.63 &  1.72  $\pm$ 0.63 \\
  $-$18.57 &    108.55 &  22.67 $\pm$  0.60 &  0.86  $\pm$ 0.64 &  1.72  $\pm$ 0.63 \\
~ & & & & \\ [-1.8ex]		      		           		         
  $-$10.99 &  $-$63.23 &  22.73 $\pm$  0.63 &  0.63  $\pm$ 0.68 &  1.15  $\pm$ 0.68 \\	    
 $-$104.57 & $-$106.84 &  22.75 $\pm$  0.62 &  0.41  $\pm$ 0.69 &  1.03  $\pm$ 0.68 \\	    
    100.28 &     56.20 &  22.75 $\pm$  0.63 &  0.57  $\pm$ 0.68 &  1.32  $\pm$ 0.67 \\
    140.89 &    114.79 &  22.77 $\pm$  0.63 &  0.74  $\pm$ 0.68 &  1.39  $\pm$ 0.67 \\
     27.38 & $-$104.95 &  22.81 $\pm$  0.64 &  0.96  $\pm$ 0.68 &  1.68  $\pm$ 0.67 
 \\ [0.8ex] \hline
\multicolumn{5}{c}{\ } \\ [-1.8ex]
\end{tabular*}
\smallskip
\parbox{15.5cm}{
{\small
\noindent 
{\sl Note to Table \ref{t:NTTphot}}: The first two columns give the positional 
offsets of the GC candidates from the centre of NGC~1316 in RA and DEC, 
respectively.}}
\end{table*}

\end{document}